\documentclass[10pt,twocolumn,floatfix,altaffilletter,superscriptaddress,tightenlines,showpacs,showkeys,preprintnumbers,nofootinbib]{revtex4-2}
\pdfoutput=1
\usepackage[colorlinks=true,citecolor=blue,linkcolor=blue,breaklinks=true,linktoc=none]{hyperref}
\usepackage{amsmath,amssymb}
\usepackage{epsfig} 
\usepackage{graphicx}
\usepackage{url}
\usepackage{color}
\usepackage{multirow}
\usepackage{placeins}
\usepackage{float}
\usepackage[dvipsnames]{xcolor}
 \usepackage{booktabs} 
\usepackage{braket}
\usepackage{float}
\usepackage{slashed,tabularray}
\hypersetup{
  colorlinks=true,
  linkcolor=red,
  filecolor=magenta,   
  urlcolor=blue,
  citecolor=blue,
}

\hypersetup{colorlinks,linkcolor={blue},citecolor={teal},urlcolor={violet}}

\allowdisplaybreaks

\bibpunct{[}{]}{,}{n}{}{,}

\def\beq{\begin{equation}}
\def\eeq{\end{equation}}
\def\bea{\begin{eqnarray}}
\def\eea{\end{eqnarray}}
\makeatletter
\@addtoreset{equation}{section}

\newcommand{\INFN}{INFN - Sezione di Napoli, Complesso Universitario Monte S. Angelo, I-80126 Napoli, Italy}

\newcommand{\SSM}{Scuola Superiore Meridionale, Università degli studi di Napoli ``Federico II'', Largo San Marcellino 10, 80138 Napoli, Italy}
\newcommand{\NAN}{Department of Physics and Institute of Theoretical Physics,
Nanjing Normal University, Nanjing, 210023, China}
\begin{document}

\title{Bayesian Forecasts on Cosmic Superstring Searches with LISA}
%  \author{Marco Chianese}
% \email{marco.chianese@unina.it}
% \affiliation{\UNINA}
%  \affiliation{\INFN}
%  \author{Satyabrata Datta}
%   \email{amisatyabrata703@gmail.com}
%   \affiliation{Department of Physics and Institute of Theoretical Physics,
%Nanjing Normal University, Nanjing, 210023, China}
\author{Satyabrata Datta}
    \email{amisatyabrata703@gmail.com}
    \affiliation{\NAN}
 \author{Rome Samanta}
  \email{samanta@na.infn.it}
  \affiliation{\SSM}
  \affiliation{\INFN}
% \author{Ninetta Saviano}
% \email{nsaviano@na.infn.it}
% \affiliation{\INFN}
%\affiliation{\SSM}

\begin{abstract}
Cosmic superstrings are well-motivated early-Universe sources of stochastic
gravitational waves, with phenomenology controlled by the string tension
\(G\mu\) and the intercommutation probability \(P\). We study whether LISA can
reconstruct these parameters and distinguish different superstring signal
models in the presence of instrumental noise and astrophysical foregrounds.
We consider two phenomenological models. In Model I, reduced intercommutation
acts only as an amplitude enhancement of a cusp-dominated spectrum,
\(\Omega_{\rm SS}^{\rm I}=P^{-\beta}\Omega_{\rm cusp}\). In Model II, the
signal is a cusp--kink mixture,
\(\Omega_{\rm SS}^{\rm II}
=P^{-\beta}[p_c\Omega_{\rm cusp}+(1-p_c)\Omega_{\rm kink}]\), so that \(P\)
controls both amplitude and spectral shape. Using simulated LISA data, we
perform Bayesian inference with instrumental-noise uncertainties, unresolved
extragalactic compact-binary backgrounds, and a flexible Galactic
double-white-dwarf foreground. We compare the models using Bayesian evidences
and map the posterior geometry in the \((G\mu,P)\) plane using marginalized
widths, correlations, covariance anisotropy, principal eigenvalues, posterior
area, and reconstruction bias. We find that reduced intercommutation
generically produces strong \(G\mu\)--\(P\) correlations, because the data often
constrain an amplitude-like parameter combination. However, when the cusp--kink
spectral difference lies in the LISA band and the signal is sufficiently loud,
Model II can be favored and the posterior can retain genuine shape information.
As an optimistic foreground scenario, we also compute the Bayes factor using a
reduced tanh Galactic foreground template, finding improved model
discrimination when the foreground shape is constrained. Our results clarify
when LISA can move beyond detecting a cosmic-superstring background to
reconstructing its microscopic parameters and discriminating between competing
superstring phenomenologies.
\end{abstract}

\maketitle
\tableofcontents
\section{Introduction}
\label{sec:introduction}

A stochastic gravitational-wave background (SGWB) \cite{Christensen:2018iqi,Romano:2016dpx,Renzini:2022alw,LISACosmologyWorkingGroup:2022jok} is not only a target for
detection, but also a test of interpretation. Many early-Universe mechanisms can
produce broad, nearly featureless spectra over the limited frequency interval
accessible to a given detector. A detection in a single band therefore does not
automatically identify the underlying source. The central question is instead
whether the data contain enough spectral information to distinguish between
different physical mechanisms once instrumental noise and astrophysical
foregrounds are included.

This question is particularly relevant for the Laser Interferometer Space
Antenna (LISA) \cite{lisa,Armano:2018kix}. Operating in the millihertz band,
LISA will probe frequencies complementary to pulsar timing arrays and
ground-based interferometers, and will be sensitive to stochastic backgrounds
from both astrophysical and cosmological sources. The recent PTA evidence for a
nanohertz common-spectrum process has further emphasized the importance of
broad-band SGWB searches, while leaving open the physical origin of the observed
signal \cite{ng1,ng2,ng3,ng4,ng5}. In this context, LISA provides a crucial
opportunity not only to detect a cosmological stochastic component, but also to
ask whether its microscopic origin can be inferred.

In this work we focus on cosmic superstrings \cite{Sarangi:2002yt,Copeland:2003bj,Jones:2003da,Dvali:2003zj,Sakellariadou:2004wq,Jackson:2004zg,Polchinski:2004ia,Davis:2005dd,Chernoff:2007pd,Damour:2004kw,Hanany:2005bc,Ellis:2023tsl,Datta:2024bqp,Ghoshal:2025tlk,Revello:2024gwa,Raidal:2026cpb,Avgoustidis:2025svu,Marfatia:2023fvh}. These are macroscopic
one-dimensional objects that can arise in string-theoretic early-Universe
scenarios, for example in brane-inflation constructions \cite{Sarangi:2002yt,Jones:2003da,Dvali:2003zj,Copeland:2003bj,Polchinski:2004ia}. Although their
microscopic origin differs from that of ordinary field-theory cosmic strings,
their cosmological evolution can share many of the same large-scale features:
strings stretch with the expansion, enter a scaling regime, reconnect, form
loops, and radiate gravitational waves \cite{Sakellariadou:2004wq,Damour:2004kw,Avelino:2012qy}. A key difference is that the
intercommutation probability need not be close to unity. Whereas ordinary
gauge-theory strings typically reconnect with probability \(P\simeq1\), cosmic
superstrings can have reduced reconnection probability, leading to a denser
network and an enhanced loop population \cite{Jackson:2004zg,Sakellariadou:2004wq,Avgoustidis:2005nv,Avgoustidis:2007aa}.

The phenomenology of realistic superstring networks can be considerably richer
than that of a single field-theory string species. Networks may contain
fundamental strings, D strings, and \((p,q)\) bound states, with
species-dependent tensions and reconnection probabilities, and may also include
junctions \cite{Copeland:2003bj,Dvali:2003zj,Jackson:2004zg,Tye:2005fn,
Copeland:2006eh,Avgoustidis:2007aa}. A full treatment of this structure is beyond the scope of the present
work. Instead, we use a reduced two-parameter description designed to isolate a
specific inference question: can LISA separately reconstruct the string tension
\(G\mu\) and the intercommutation probability \(P\), and can it distinguish
between different ways in which reduced intercommutation affects the observed
spectrum?

We consider two phenomenological superstring models. In Model I, the effect of
reduced intercommutation is taken to be an amplitude rescaling of a
cusp-dominated cosmic-string spectrum,
\[
    \Omega_{\rm SS}^{\rm I}(f;G\mu,P)
    =
    P^{-\beta}\Omega_{\rm cusp}(f;G\mu).
\]
This model captures the standard expectation that smaller \(P\) increases the
number density of strings and loops, thereby enhancing the stochastic
background. It also provides a useful baseline for parameter reconstruction:
if the spectrum locally scales as \(\Omega_{\rm cusp}\propto(G\mu)^q\), the data
primarily constrain the amplitude combination
\[
    q\log_{10}G\mu-\beta\log_{10}P\simeq {\rm const.},
\]
so that \(G\mu\) and \(P\) are expected to be strongly correlated. Model I was also considered by the NANOGrav collaboration as a possible interpretation of the observed nanohertz
common-spectrum process \cite{ng5}.

Model II extends this amplitude-only picture by allowing the small-scale
structure of loops to change with \(P\). 
For compactness, we write
\(\Omega_{\rm cusp,kink}=\Omega_{\rm cusp,kink}(f;G\mu)\) and
\(p_c=p_c(P)\). Then
\[
\begin{aligned}
\Omega_{\rm SS}^{\rm II}
=
P^{-\beta}
\Big[
    p_c\,\Omega_{\rm cusp}
    +(1-p_c)\,\Omega_{\rm kink}
\Big].
\end{aligned}
\]
where \(p_c(P)\equiv p_{\rm cusp}(P)\) is the cusp fraction. In this model,
\(P\) controls not only the overall amplitude but also the relative weight of
cusp- and kink-dominated emission. We use the smooth benchmark form
\[
    p_c(P)
    =
    \frac{P^\eta}{P^\eta+P_0^\eta},
\]
with \(P_0=10^{-2}\) and \(\eta=1\). Thus, for \(P\gg P_0\), Model II reduces
to the cusp-dominated amplitude-rescaled spectrum of Model I, while for
\(P\ll P_0\) it becomes increasingly kink dominated. This construction allows
us to test whether LISA can identify shape information associated with the
cusp--kink mixture, rather than merely measuring a single amplitude
combination.

The distinction between these two models is not guaranteed to be observable.
If the LISA band samples a frequency range in which the cusp and kink spectra
are nearly degenerate, or if the cosmological signal is partially absorbed by
foreground freedom, Model II may be statistically indistinguishable from
Model I even when it is the true signal model. Conversely, if the cusp--kink
shape difference lies in the sensitive part of the LISA band and the signal is
sufficiently loud, the data can favor the shape-extended model over the
amplitude-only model. The problem is therefore one of model identifiability, not
only parameter estimation.

Foreground modeling is central to this reconstruction problem. In the LISA
band, the unresolved Galactic population of double-white-dwarf binaries is
expected to generate a strong confusion foreground
\cite{Adams:2013qma,Boileau:2020rpg,Boileau:2021sni,Korol:2021pun,
Korol:2020lpq,Liu:2023qap}. In addition, unresolved extragalactic compact
binaries contribute stochastic components with spectra that can overlap the
cosmological signal
\cite{Phinney:2001di,Regimbau:2011rp,Babak:2023lro,Lehoucq:2023zlt}. The
observable LISA background is therefore not a superstring spectrum in isolation,
but a superposition of instrumental noise, astrophysical foregrounds, and a
possible cosmological contribution.

We account for this by analyzing the superstring signal jointly with
instrumental-noise parameters, unresolved extragalactic compact-binary
backgrounds, and a Galactic double-white-dwarf foreground. Our baseline analysis
uses a flexible Galactic template, allowing both its amplitude and spectral
shape to vary. This provides a conservative inference setup: the superstring
parameters are reconstructed only after marginalizing over foreground and noise
uncertainties, rather than in an idealized foreground-free limit. As a
complementary optimistic case, we also consider a reduced tanh Galactic
foreground template \cite{Caprini:2024hue,Blanco-Pillado:2024aca,Karnesis:2021tsh,Samanta:2025jec}, in which the foreground shape is fixed and only its overall
normalization is varied. This comparison allows us to assess how improved prior
knowledge of the Galactic foreground would affect superstring model
discrimination.

\begin{figure*}
    \centering
    \includegraphics[width=.8\linewidth]{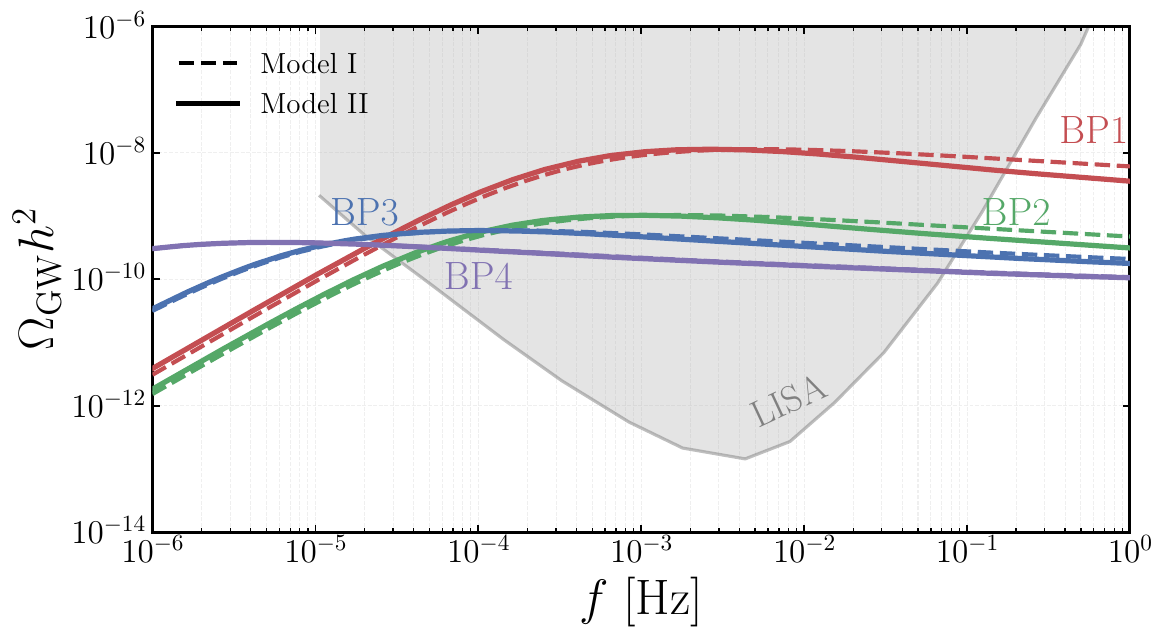} 
   \caption{
Benchmark cosmic-superstring spectra used in the subsequent reconstruction
analysis. The dashed curves show Model I, the amplitude-only cusp model,
\(\Omega_{\rm SS}^{\rm I}=P^{-\beta}\Omega_{\rm cusp}\), while the solid curves
show Model II, the shape-extended cusp--kink mixture,
\(\Omega_{\rm SS}^{\rm II}
=P^{-\beta}[p_c\Omega_{\rm cusp}+(1-p_c)\Omega_{\rm kink}]\).
The four benchmark points BP1--BP4 sample representative regions of the
\((G\mu,P)\) plane and are marked consistently in the Bayes-factor and posterior
diagnostic maps below. The gray shaded region indicates the approximate LISA
sensitivity band. The comparison illustrates where the two superstring models
produce nearly identical spectra and where the cusp--kink mixture induces an
observable shape difference in the LISA band.
}
\label{fig:bp_superstring_spectra}
\end{figure*}

To connect the signal models with the reconstruction problem,
Fig.~\ref{fig:bp_superstring_spectra} shows representative benchmark spectra
for the two superstring descriptions. These benchmark points sample different
regions of the \((G\mu,P)\) plane and illustrate when the amplitude-only and
cusp--kink spectra are nearly degenerate, and when their LISA-band shapes differ
appreciably. The same benchmark points are marked consistently in the
Bayes-factor maps, posterior-diagnostic maps, and explicit posterior
reconstructions below.
The goal of this paper is to construct an inference-level reconstruction
landscape for cosmic superstrings in LISA. We first compare Model I and Model II
using Bayesian evidences, identifying where the data can distinguish an
amplitude-only reduced-intercommutation effect from a genuine cusp--kink shape
modification. We then study the posterior geometry within each model using
marginalized widths, parameter correlations, covariance anisotropy, principal
directions, localization area, and reconstruction bias. This allows us to
separate four physically different outcomes: non-detection, detection without
model discrimination, compact but correlated reconstruction of a parameter
combination, and genuine two-parameter reconstruction of \((G\mu,P)\). We refer the readers to Refs.~\cite{Caprini:2019pxz,Flauger:2020qyi,Caprini:2024hue,Blanco-Pillado:2024aca,LISACosmologyWorkingGroup:2025vdz,Gowling:2021gcy,Giese:2021dnw,Gowling:2022pzb,Samanta:2025jec,Chen:2023zkb,Guan:2025idx,Kume:2024xvh,Dimitriou:2025bvq,Dimitriou:2026agw,LISACosmologyWorkingGroup:2024hsc,Datta:2026hka} for recent reconstruction works of various cosmological sources in the LISA band.

The paper is organized as follows. In Sec.~\ref{sec:gw_superstrings}, we review
the gravitational-wave spectra used for the two cosmic-superstring models. In
Sec.~\ref{sec:lisa_inference_superstrings}, we describe the LISA data model,
including instrumental noise and astrophysical foregrounds. In
Sec.~\ref{sec:posterior_diagnostics_superstrings}, we define the posterior
diagnostics used to interpret reconstruction quality. In
Sec.~\ref{sec:results_superstrings}, we present the Bayes-factor maps,
posterior-covariance diagnostics, bias maps, and benchmark posterior
distributions. We summarize our conclusions in Sec.~\ref{sec:conclusions}.

\section{Gravitational waves from cosmic superstrings}
\label{sec:gw_superstrings}

Cosmic superstrings are macroscopic one-dimensional objects that can be formed
in string-theoretic scenarios of the early Universe, for example after brane
inflation or other symmetry-breaking-like transitions in the compactified
theory \cite{Sarangi:2002yt,Jones:2003da,Dvali:2003zj,Copeland:2003bj,Polchinski:2004ia}. Although their microscopic origin is different from that of ordinary
field-theory cosmic strings, their cosmological evolution can be analogous:
after formation, they can stretch with the expansion of the Universe, approach a
scaling network, self-intersect or reconnect, form closed loops, and lose energy
through gravitational-wave emission \cite{Sakellariadou:2004wq,Damour:2004kw,Avelino:2012qy}. They therefore provide a string-theoretic
realization of cosmic-string gravitational-wave phenomenology.

Cosmic superstrings differ from ordinary field-theory strings in several
important respects. Field-theory strings typically reconnect with probability
close to unity, whereas cosmic superstrings can have reduced intercommutation
probabilities. This can increase the density of the long-string network and
enhance the abundance of radiating loops. In addition, realistic superstring
networks may contain several string species, including F strings, D strings, and
$(p,q)$ bound states, with species-dependent tensions, reconnection
probabilities, and junctions
\cite{Jackson:2004zg,Copeland:2003bj,Dvali:2003zj,Sarangi:2002yt,
Polchinski:2004ia,Avgoustidis:2005nv,Tye:2005fn}. These properties can modify
both the normalization and, in more general cases, the shape of the stochastic
gravitational-wave background.

The possibility that cosmic strings or superstrings could contribute to a
nanohertz stochastic background has received renewed interest in light of recent
pulsar-timing-array evidence for a common-spectrum process, see, e.g., Refs.\cite{ng5,Ellis:2023tsl,Datta:2024bqp}. While the
interpretation of the PTA signal remains open, this has sharpened the broader
question of whether future gravitational-wave observations can distinguish
standard cosmic strings from superstring-motivated extensions. In this work we
address this question at the level of LISA reconstruction: can the
intercommutation probability be inferred independently, or does it remain
degenerate with the string tension?

We adopt a phenomenological description designed to isolate two leading effects
of cosmic-superstring microphysics. First, reduced intercommutation can enhance
the loop number density, producing an approximately amplitude-like
rescaling of the stochastic background \cite{ng5,Ellis:2023tsl,Datta:2024bqp}. Second, superstring dynamics may modify
the relative importance of cusp- and kink-dominated loop emission \cite{OCallaghan:2010mtk,Binetruy:2009vt,Damour:2001bk,Olmez:2010bi,Matsui:2020hzi}. The first
effect is expected to be strongly degenerate with the string tension, while the
second can introduce additional shape dependence through the harmonic power
spectrum.

We begin with the standard loop contribution. A loop formed at time $t_i$ with
initial length
\begin{equation}
    l_i=\alpha t_i
\end{equation}
shrinks through gravitational-wave emission as \cite{Vilenkin:1981bx,Vachaspati:1984gt}
\begin{equation}
    l(t)
    =
    l_i
    -
    \Gamma_{\rm GW}G\mu\,(t-t_i),
    \label{eq:super_loop_evolution}
\end{equation}
where $\Gamma_{\rm GW}\simeq 50$ \cite{Vilenkin:1981bx,Vachaspati:1984gt}, $\alpha\simeq 0.1$ \cite{Blanco-Pillado:2013qja,Blanco-Pillado:2017oxo}, and $G\mu$ is the
dimensionless string tension. The loop emits into normal modes labelled by
$j=1,2,\dots,j_{\rm max}$, with emitted frequency
\begin{equation}
    f_{\rm em}^{(j)}
    =
    \frac{2j}{l_j(t)}.
\end{equation}
The observed frequency today is related to the emitted frequency by
\begin{equation}
    f
    =
    \frac{a(t)}{a(t_0)}
    f_{\rm em}^{(j)},
\end{equation}
or equivalently
\begin{equation}
    l_j(t)
    =
    \frac{2j}{f}
    \frac{a(t)}{a(t_0)} .
\end{equation}

For an ordinary unit-reconnection string network, the present-day
gravitational-wave energy density from loops can be written schematically as \cite{Blanco-Pillado:2013qja,Blanco-Pillado:2017oxo}
\begin{equation}
\Omega_{\rm NG}(f;G\mu,\delta)
=
\sum_{j=1}^{j_{\rm max}}
\frac{2j\,G\mu^2\,\Gamma_j(\delta)}
     {f\,\rho_c}
\int_{t_F}^{t_0} dt\,
\left[\frac{a(t)}{a(t_0)}\right]^5
n(t,l_j),
\label{eq:omega_ng_superstring}
\end{equation}
where $\rho_c$ is the critical density, $n(t,l_j)$ is the loop number density
evaluated at the loop length contributing to the observed frequency $f$ \cite{Martins:1996jp,Martins:2000cs,Sousa:2013aaa,Auclair:2019wcv,Datta:2024bqp}, and
$\Gamma_j$ is the power emitted into the $j$th harmonic. We use
\begin{equation}
    \Gamma_j(\delta)
    =
    \frac{\Gamma_{\rm GW}\,j^{-\delta}}
         {\zeta(\delta)} ,
    \label{eq:gamma_j_delta}
\end{equation}
with $\delta=4/3$ for cusp-dominated emission and $\delta=5/3$ for
kink-dominated emission
\cite{Damour:2001bk}. The
normalization is chosen such that
\begin{equation}
    \sum_{j=1}^{\infty}\Gamma_j=\Gamma_{\rm GW}.
\end{equation}
Although the total emitted power is fixed by $\Gamma_{\rm GW}$, the observed
spectrum depends on the full harmonic distribution because the sum contains both
the explicit harmonic weight and the loop density evaluated at
$l_j\propto j$. Therefore, cusp- and kink-dominated emission can produce
different gravitational-wave spectra even for the same total loop power.

We define
\begin{equation}
    \Omega_{\rm cusp}(f;G\mu)
    \equiv
    \Omega_{\rm NG}\!\left(f;G\mu,\delta=\frac{4}{3}\right),
\end{equation}
and
\begin{equation}
    \Omega_{\rm kink}(f;G\mu)
    \equiv
    \Omega_{\rm NG}\!\left(f;G\mu,\delta=\frac{5}{3}\right).
\end{equation}

\subsection{Model I: amplitude-only reduced intercommutation}

The simplest way to include reduced intercommutation is to rescale the loop
number density by a power of the intercommutation probability $P$,
\begin{equation}
    n(t,l;P)
    =
    P^{-\beta}
    n(t,l;P=1),
    \label{eq:p_beta_loop_density}
\end{equation}
where $\beta$ parametrizes the strength of the network-density enhancement \cite{Sakellariadou:2004wq,Damour:2004kw,Avelino:2012qy,ng5,Avgoustidis:2005nv,Avgoustidis:2007aa}. In
the first model considered in this work, we fix
\begin{equation}
    \beta=1,
\end{equation}
corresponding to the commonly used prescription $n\rightarrow P^{-1}n$ \cite{ng5,Ellis:2023tsl,Datta:2024bqp} (see Ref.\cite{Avgoustidis:2005nv} for a different scaling relation). At the
level of the gravitational-wave spectrum, this gives
\begin{equation}
    \Omega_{\rm SS}^{\rm I}(f;G\mu,P)
    =
    P^{-1}
    \Omega_{\rm NG}(f;G\mu,\delta_{\rm c}),
    \label{eq:omega_superstring_model1}
\end{equation}
where $\delta_{\rm c}=4/3$ is used as the fiducial cusp-dominated harmonic
spectrum.

This model captures the leading expectation that smaller $P$ produces a denser
network and therefore a larger stochastic background. However, since $P$ enters
only as an overall normalization, it is expected to be degenerate with
$G\mu$---at least in the high-frequency part of the spectrum. If, over the frequency range probed by the detector, the spectrum scales
locally as
\begin{equation}
    \Omega_{\rm NG}(f;G\mu)\propto (G\mu)^q ,
\end{equation}
then the data primarily constrain the effective amplitude combination
\begin{equation}
    (G\mu)^q P^{-1}
    \simeq
    {\rm const.}
    \label{eq:gmu_p_amp_degeneracy}
\end{equation}
Thus Model I provides a baseline case in which the intercommutation probability
is mainly an amplitude parameter.

\subsection{Model II: cusp--kink mixture with a transition scale}

We next consider a shape-extended model in which the intercommutation
probability affects not only the overall network density, but also the relative
importance of cusp- and kink-dominated emission. The motivation is that cusp
formation in cosmic-superstring networks can be suppressed by motion in compact
extra dimensions, while junctions and small-scale structure can enhance the
importance of kink-like emission
\cite{OCallaghan:2010mtk,Binetruy:2009vt,Damour:2001bk,Olmez:2010bi,Matsui:2020hzi}. We model this effect
phenomenologically as
\begin{equation}
\begin{aligned}
    \Omega_{\rm SS}^{\rm II}(f;G\mu,P)
    =
    P^{-\beta}
    \Big[
    &p_{\rm cusp}(P)\,
    \Omega_{\rm cusp}(f;G\mu)
    \\
    &+
    \big(1-p_{\rm cusp}(P)\big)
    \Omega_{\rm kink}(f;G\mu)
    \Big].
\end{aligned}
\label{eq:omega_superstring_model2}
\end{equation}
Equivalently, the harmonic power spectrum is
\begin{equation}
    \Gamma_j(P)
    =
    \Gamma_{\rm GW}
    \left[
    p_{\rm cusp}(P)
    \frac{j^{-4/3}}{\zeta(4/3)}
    +
    \left(1-p_{\rm cusp}(P)\right)
    \frac{j^{-5/3}}{\zeta(5/3)}
    \right].
    \label{eq:gamma_j_superstring_model2}
\end{equation}
Note that this mixture is not exactly equivalent to a single effective harmonic
power law of the form \(j^{-\delta_{\rm eff}}\), with
\(\delta_{\rm eff}\) treated as a constant. To see this, consider
\begin{equation}
    \Gamma_j
    =
    A j^{-4/3}
    +
    B j^{-5/3},
\end{equation}
where \(A\) and \(B\) are the cusp and kink weights. A local effective index can
be defined as
\begin{equation}
    \delta_{\rm eff}(j)
    =
    -\frac{d\ln\Gamma_j}{d\ln j}.
\end{equation}
For the mixed spectrum, this gives
\begin{equation}
    \delta_{\rm eff}(j)
    =
    \frac{
    \frac{4}{3}A j^{-4/3}
    +
    \frac{5}{3}B j^{-5/3}
    }{
    A j^{-4/3}
    +
    B j^{-5/3}
    } .
\end{equation}
Thus the effective index depends on both the relative cusp--kink weights and the
harmonic number \(j\). It approaches \(4/3\) in the cusp-dominated limit and
\(5/3\) in the kink-dominated limit, but in the mixed regime it is not a
constant. Moreover, the stochastic background depends on the full harmonic sum
and on the loop density evaluated at the harmonic-dependent length \(l_j\), so
the mixture cannot in general be reduced to a single fixed
\(j^{-\delta_{\rm eff}}\) spectrum.

For the cusp fraction we use the transition ansatz
\begin{equation}
    p_{\rm cusp}(P)
    =
    \frac{P^\eta}{P^\eta+P_0^\eta}.
    \label{eq:pcusp_transition}
\end{equation}
Here $P_0$ sets the transition scale and $\eta$ controls the sharpness of the
transition. By construction,
\begin{equation}
    p_{\rm cusp}(P_0)=\frac{1}{2}.
\end{equation}
For $P\gg P_0$, the spectrum is cusp dominated, while for $P\ll P_0$ it is
kink dominated. In the present analysis, we fix
\begin{equation}
    P_0=10^{-2},
    \qquad
    \eta=1,
\end{equation}
so that the cusp-to-kink transition is smooth and centered around
$P\simeq10^{-2}$.

This parametrization makes explicit that the shape sensitivity to $P$ is
localized near the transition region $P\sim P_0$. Away from this region, the
model approaches an amplitude-rescaled single-shape spectrum:
\begin{equation}
    P\gg P_0:
    \quad
    \Omega_{\rm GW}^{\rm II}
    \simeq
    P^{-\beta}\Omega_{\rm cusp},
\end{equation}
and
\begin{equation}
    P\ll P_0:
    \quad
    \Omega_{\rm GW}^{\rm II}
    \simeq
    P^{-\beta}\Omega_{\rm kink}.
\end{equation}
Thus, outside the transition region, $P$ again behaves mainly as an amplitude
parameter and is expected to remain strongly degenerate with $G\mu$. Model II
can weaken this degeneracy only when the detector is sensitive to the spectral
difference between the cusp- and kink-dominated components within the observed
frequency band.

In the reconstruction analysis, we use the two-dimensional signal-parameter
space
\begin{equation}
    \theta
    =
    \left(
    \log_{10}G\mu,\,
    \log_{10}P
    \right),
\end{equation}
fixing $\beta=1$ in each model and fixing $(P_0,\eta)=(10^{-2},1)$ in Model II.
Comparing Model I with Model II allows us to determine whether LISA can
reconstruct the intercommutation probability only through an effective
amplitude combination, or whether superstring-induced changes in the
loop-emission structure leave a measurable spectral imprint.
\section{LISA data model, foregrounds, and inference setup}
\label{sec:lisa_inference_superstrings}

We use the same LISA inference framework as in our domain wall and metastable-string analysis, with the cosmological signal replaced by the cosmic-superstring spectrum \cite{Datta:2026fav,Datta:2026ffs,Datta:2026hka}. Here we summarize the ingredients relevant for the present reconstruction; further
details of the detector response, foreground modeling, likelihood construction,
and mock-data generation can be found in Refs.~\cite{Datta:2026fav,Datta:2026ffs,Datta:2026hka}.

\subsection{LISA configuration, TDI observables, and instrumental noise}
\label{subsec:lisa_tdi_noise_superstrings}

The analysis is performed in the approximately noise-orthogonal time-delay
interferometry basis \cite{Tinto:2004wu,McNamara:2008zz,Armano:2018kix,Caprini:2019pxz,Flauger:2020qyi,Caprini:2024hue}
\begin{equation}
    \tilde d_a(f),
    \qquad
    a\in\{A,E,T\}.
\end{equation}
The \(A\) and \(E\) channels provide the dominant sensitivity to an isotropic
stochastic gravitational-wave background, while the \(T\) channel is treated as
an approximate null channel at low frequencies and is used to help constrain the
instrumental noise.

For each channel, the measured power spectral density is modeled as \cite{Tinto:2004wu,McNamara:2008zz,Armano:2018kix,Caprini:2019pxz,Flauger:2020qyi,Caprini:2024hue}
\begin{equation}
    P_a(f;\boldsymbol{\theta})
    =
    N_a(f;\boldsymbol{\theta})
    +
    S_a(f;\boldsymbol{\theta}),
    \qquad
    a\in\{A,E,T\},
\end{equation}
where \(N_a\) is the instrumental-noise contribution and \(S_a\) is the
gravitational-wave contribution. The signal spectrum is related to the
underlying energy-density spectrum by
\begin{equation}
    S_a(f;\boldsymbol{\theta})
    =
    \frac{3H_0^2}{4\pi^2}
    \frac{\Omega_{\rm GW}(f;\boldsymbol{\theta})}{f^3}
    \mathcal R_a(f),
\end{equation}
where \(\mathcal R_a(f)\) is the sky- and polarization-averaged LISA response
function \cite{Cornish:2001bb,Smith:2019wny}. The leading instrumental-noise amplitudes, associated with residual
test-mass acceleration noise and optical metrology noise, are treated as
nuisance parameters and marginalized over in the inference.

\subsection{Astrophysical foregrounds and cosmic-superstring signal model}
\label{subsec:lisa_foregrounds_superstrings}

A cosmological stochastic background in the LISA band must be inferred in the
presence of unresolved astrophysical foregrounds. We model the total
gravitational-wave energy density as
\begin{equation}
    \Omega_{\rm GW}(f;\boldsymbol{\theta})
    =
    \Omega_{\rm DWD}(f;\boldsymbol{\theta})
    +
    \Omega_{\rm ast}(f;\boldsymbol{\theta})
    +
    \Omega_{\rm SS}(f;G\mu,P),
    \label{eq:total_omega_superstrings}
\end{equation}
where \(\Omega_{\rm DWD}\) is the unresolved Galactic double-white-dwarf
foreground, \(\Omega_{\rm ast}\) is the unresolved extragalactic compact-binary
foreground, and \(\Omega_{\rm SS}\) is the cosmic-superstring contribution.

For the Galactic foreground, we adopt, as our baseline, the same flexible
phenomenological template used in Refs.~\cite{Korol:2021pun,Korol:2020lpq,Liu:2023qap,Chen:2023zkb,Datta:2026ffs,Datta:2026fav},
\begin{equation}
    \Omega_{\rm DWD}(f)
    =
    \frac{
    A_1\left(f/f_{\rm ref}\right)^{\alpha_1}
    }{
    1+
    A_2\left(f/f_{\rm ref}\right)^{\alpha_2}
    },
    \label{eq:dwd_model_superstrings}
\end{equation}
where \(A_1,A_2,\alpha_1,\alpha_2\) control the normalization and spectral
shape of the unresolved Galactic foreground. This template is deliberately
conservative: by allowing both the amplitude and shape of the foreground to vary,
it can absorb broad spectral curvature in the mHz band and therefore weakens the
apparent reconstruction power for a cosmological signal.

We later compare this conservative choice with a reduced tanh Galactic
foreground template \cite{Caprini:2024hue,Blanco-Pillado:2024aca,Karnesis:2021tsh,Samanta:2025jec}, in which the spectral shape is fixed by the observation
time and only the overall normalization is allowed to vary. The tanh template is
less flexible and therefore less able to mimic broad cosmological spectral
features. It provides a more optimistic foreground scenario and allows us to
assess how much the reconstruction of \((G\mu,P)\) improves when the Galactic
foreground shape is assumed to be better characterized.

The unresolved extragalactic foreground is modeled as a power law \cite{Phinney:2001di,Regimbau:2011rp,Babak:2023lro,Lehoucq:2023zlt},
\begin{equation}
    \Omega_{\rm ast}(f)
    =
    \Omega_{\rm ast}
    \left(
    \frac{f}{f_{\rm ref}}
    \right)^{\varepsilon}.
    \label{eq:astro_power_law_superstrings}
\end{equation}
Over the LISA frequency range considered here, this provides an effective
description of unresolved compact-binary backgrounds.

The baseline parameter vector is therefore
\begin{equation}
    \boldsymbol{\theta}
    =
    \left\{
    N_{\rm acc},
    \delta x,
    A_1,
    \alpha_1,
    A_2,
    \alpha_2,
    \Omega_{\rm ast},
    \varepsilon,
    G\mu,
    P
    \right\}.
    \label{eq:theta_superstring_lisa}
\end{equation}
Here \(N_{\rm acc}\) and \(\delta x\) describe the leading instrumental-noise
amplitudes, \((A_1,\alpha_1,A_2,\alpha_2)\) describe the flexible Galactic
foreground template, \((\Omega_{\rm ast},\varepsilon)\) describe the
extragalactic foreground, and \((G\mu,P)\) are the cosmic-superstring parameters
of interest. In the signal sector, we sample
\begin{equation}
    \theta_{\rm sig}
    =
    \left(
    \log_{10}G\mu,\log_{10}P
    \right).
\end{equation}
The prior ranges used in the baseline analysis are shown in
Table~\ref{tab:lisa_parameters}.

\begin{table}[htbp]
\centering
\begin{tblr}{
    hlines,
    vlines,
    row{1} = {bg=gray7, fg=white, font=\bfseries},
    column{1} = {bg=gray9},
    cell{1}{1} = {bg=gray7, fg=white},
}
\textbf{Parameter} & \textbf{Fiducial value} & \textbf{Uniform prior} \\
\(\log_{10}(N_{\rm acc})\)      & \(-14.523\) & \((-16.0,-13.7)\) \\
\(\log_{10}(\delta x)\)         & \(-11.097\) & \((-13.0,-10.7)\) \\
\(\log_{10}(A_1)\)              & \(-15.4\)   & \((-18.0,-5.0)\) \\
\(\alpha_1\)                    & \(-5.7\)    & \((-15.0,-3.0)\) \\
\(\log_{10}(A_2)\)              & \(-6.32\)   & \((-10.0,5.0)\) \\
\(\alpha_2\)                    & \(-6.2\)    & \((-10.0,-1.0)\) \\
\(\log_{10}(\Omega_{\rm ast})\) & \(-11.0\)   & \((-15.0,-8.0)\) \\
\(\varepsilon\)                 & \(0.67\)    & \((0.0,1.0)\) \\
\(\log_{10}(G\mu)\)             & grid        & \((-20,-12)\) \\
\(\log_{10}(P)\)                & grid        & \((-4,0)\) \\
\end{tblr}
\caption{
Model parameters, fiducial values, and prior ranges used in the baseline
inference. The first two parameters describe the leading instrumental-noise
amplitudes, the next six describe astrophysical foregrounds, and the final two
determine the cosmic-superstring signal.
}
\label{tab:lisa_parameters}
\end{table}
\subsection{Frequency-domain likelihood and synthetic data generation}
\label{subsec:lisa_likelihood_superstrings}

The likelihood is constructed from the complex Fourier coefficients of the TDI
data \cite{Boileau:2020rpg,Romano:2016dpx,Smith:2019wny,Guan:2025idx,Datta:2026fav,Datta:2026ffs}. The observation is divided into \(N_{\rm seg}\) approximately stationary
segments of duration \(T\). For a sampling interval \(\Delta t\), the sampling
frequency is \(f_s=1/\Delta t\), and each segment contains
\begin{equation}
    N=\frac{T}{\Delta t}
\end{equation}
time samples. The discrete Fourier frequencies are
\begin{equation}
    f_k=k\Delta f,
    \qquad
    \Delta f=\frac{1}{T},
    \qquad
    k=0,1,\dots,\frac{N}{2}.
\end{equation}
For each segment \(r=1,\dots,N_{\rm seg}\) and positive-frequency bin \(f_k\),
we define
\begin{equation}
    \mathbf d_{rk}
    =
    \bigl(\tilde d_A^r,\tilde d_E^r,\tilde d_T^r\bigr)^{\rm T}_{f=f_k}.
\end{equation}

We assume that \(\mathbf d_{rk}\) is drawn from a zero-mean complex Gaussian
distribution with covariance
\begin{equation}
    C_k(\boldsymbol{\theta})
    =
    \frac{Tf_s^2}{2}\,
    {\rm diag}\!\left(P_A,P_E,P_T\right)_{f=f_k}.
\end{equation}
The diagonal form follows from the use of the approximately noise-orthogonal
\((A,E,T)\) basis. For one segment and one frequency bin, the corresponding
probability density is
\begin{equation}
    p(\mathbf d_{rk}|\boldsymbol{\theta})
    =
    \frac{1}{\pi^3\det C_k(\boldsymbol{\theta})}
    \exp\left[
    -\mathbf d_{rk}^{\dagger}
    C_k^{-1}(\boldsymbol{\theta})
    \mathbf d_{rk}
    \right].
    \label{eq:single_bin_likelihood_superstrings}
\end{equation}
Assuming statistical independence among segments and Fourier bins, the full
likelihood is
\begin{equation}
    \mathcal L(\boldsymbol{\theta})
    =
    \prod_{r=1}^{N_{\rm seg}}
    \prod_{k=1}^{N/2}
    p(\mathbf d_{rk}|\boldsymbol{\theta}) .
\end{equation}

We use the null-channel approximation
\begin{equation}
    P_A=N_A+S_A,
    \qquad
    P_E=N_E+S_E,
    \qquad
    P_T=N_T ,
\end{equation}
where the frequency and parameter dependence is implicit. Dropping additive
constants independent of \(\boldsymbol{\theta}\), the frequency-domain
log-likelihood becomes \cite{Boileau:2020rpg,Romano:2016dpx,Smith:2019wny,Guan:2025idx,Datta:2026fav,Datta:2026ffs}
\begin{widetext}
\begin{equation}
\begin{aligned}
    \ln\mathcal L(\boldsymbol{\theta})
    =
    -\sum_{r=1}^{N_{\rm seg}}
    \sum_{k=1}^{N/2}
    \Bigg\{
        &\ln\left[
        \big(N_A+S_A\big)
        \big(N_E+S_E\big)
        N_T
        \right]_{f=f_k}
        \\
        &+
        \frac{2}{Tf_s^2}
        \left[
        \frac{|\tilde d_A^r(f_k)|^2}
        {N_A(f_k;\boldsymbol{\theta})+S_A(f_k;\boldsymbol{\theta})}
        +
        \frac{|\tilde d_E^r(f_k)|^2}
        {N_E(f_k;\boldsymbol{\theta})+S_E(f_k;\boldsymbol{\theta})}
        +
        \frac{|\tilde d_T^r(f_k)|^2}
        {N_T(f_k;\boldsymbol{\theta})}
        \right]
    \Bigg\}.
\end{aligned}
\label{eq:lisa_loglike_superstrings}
\end{equation}
\end{widetext}

Synthetic datasets are generated by choosing an injected parameter vector
\(\boldsymbol{\theta}_{\rm inj}\) and drawing the TDI Fourier coefficients from
the corresponding complex Gaussian distribution,
\begin{equation}
    \mathbf d_{rk}^{\rm mock}
    \sim
    \mathcal N_{\mathbb C}
    \left(
    0,
    C_k(\boldsymbol{\theta}_{\rm inj})
    \right).
\end{equation}
The posterior distribution is
\begin{equation}
    p(\boldsymbol{\theta}|d)
    =
    \frac{
    \mathcal L(d|\boldsymbol{\theta})\pi(\boldsymbol{\theta})
    }{
    Z
    },
    \qquad
    Z=
    \int d\boldsymbol{\theta}\,
    \mathcal L(d|\boldsymbol{\theta})\pi(\boldsymbol{\theta}) .
\end{equation}
Posterior samples are used to construct reconstruction maps, marginalized
uncertainties, covariance diagnostics, and bias estimates in the
\((\log_{10}G\mu,\log_{10}P)\) plane. In the numerical implementation, the
posterior is explored with nested sampling, using \texttt{dynesty} \cite{speagle2020dynesty} interfaced
through \texttt{Bilby} \cite{Ashton:2018jfp}.
\section{Posterior diagnostics }
\label{sec:posterior_diagnostics_superstrings}
\begin{table*}[t]
\centering
\begin{tblr}{
    colspec={p{0.24\linewidth} p{0.24\linewidth} p{0.38\linewidth}},
    hlines,
    vlines,
    row{1}={bg=gray7, fg=White, font=\bfseries},
    column{1}={bg=gray7},
}
Diagnostic pattern & Reconstruction regime & Interpretation \\
\hline

1. \(\kappa_{\rm deg}\gg1\), 
\(|\rho|\simeq1\), 
small \(A_{\rm cov}\), 
small \(\lambda_\perp\)
&
Compact correlated reconstruction
&
The posterior is highly elongated but well localized. LISA measures one
combination of \(G\mu\) and \(P\) very accurately, while the remaining
uncertainty follows a correlated trade-off direction. This is not a bad
reconstruction. \\

2. \(\kappa_{\rm deg}\gg1\), 
\(|\rho|\simeq1\), 
large \(A_{\rm cov}\)
&
Broad correlated degeneracy
&
The data constrain one parameter combination, but the allowed ridge is wide.
Individual reconstruction of \(G\mu\) and \(P\) is weak. \\

3. \(\kappa_{\rm deg}\gg1\), 
\(|\rho|\ll1\), 
one large \(\sigma_i\)
&
Axis-aligned loss of identifiability
&
One parameter direction is weakly constrained, often because changing that
parameter has little effect on the LISA-band signal. \\

4. \(\kappa_{\rm deg}\sim1\), 
\(|\rho|\ll1\), 
small \(A_{\rm cov}\), 
small \(\sigma_i\)
&
Balanced two-parameter reconstruction
&
Both \(G\mu\) and \(P\) are independently localized. This is the clearest case
of genuine two-parameter reconstruction. \\

5. \(\kappa_{\rm deg}\sim1\), 
large \(A_{\rm cov}\), 
large \(\sigma_i\)
&
Broad weak reconstruction
&
The posterior is not strongly elongated, but it is still broad. Small
\(\kappa_{\rm deg}\) alone does not imply good reconstruction. \\

6. \(A_{\rm cov}\), \(\lambda_\perp\), and \(\sigma_i\) comparable to prior widths
&
Prior-dominated regime
&
The likelihood carries little information. Apparent anisotropy or isotropy is
mainly inherited from the prior geometry. \\

7. small \(A_{\rm cov}\), 
small \(\sigma_i\), 
large reconstruction bias
&
Precise but inaccurate reconstruction
&
The model selects a compact region, but not the injected one. This can indicate
foreground degeneracy, model mismatch, prior-boundary effects, or a misleading
posterior mean. \\
\end{tblr}
\caption{
Interpretation of the posterior diagnostics used in the
\((\log_{10}G\mu,\log_{10}P)\) reconstruction maps. Here
\(\rho\) measures parameter correlation in the chosen coordinates,
\(\kappa_{\rm deg}\) measures posterior anisotropy, \(A_{\rm cov}\) measures
overall localization, \(\lambda_\perp\) measures the best-constrained principal
direction, and \(\sigma_i\) are marginalized widths. Good reconstruction
requires small localization measures and accurate recovery, not simply small
\(\kappa_{\rm deg}\) or weak correlation.
}
\label{tab:superstring_diagnostic_interpretation}
\end{table*}
\begin{figure*}
    \centering
    \includegraphics[width=0.57\linewidth]{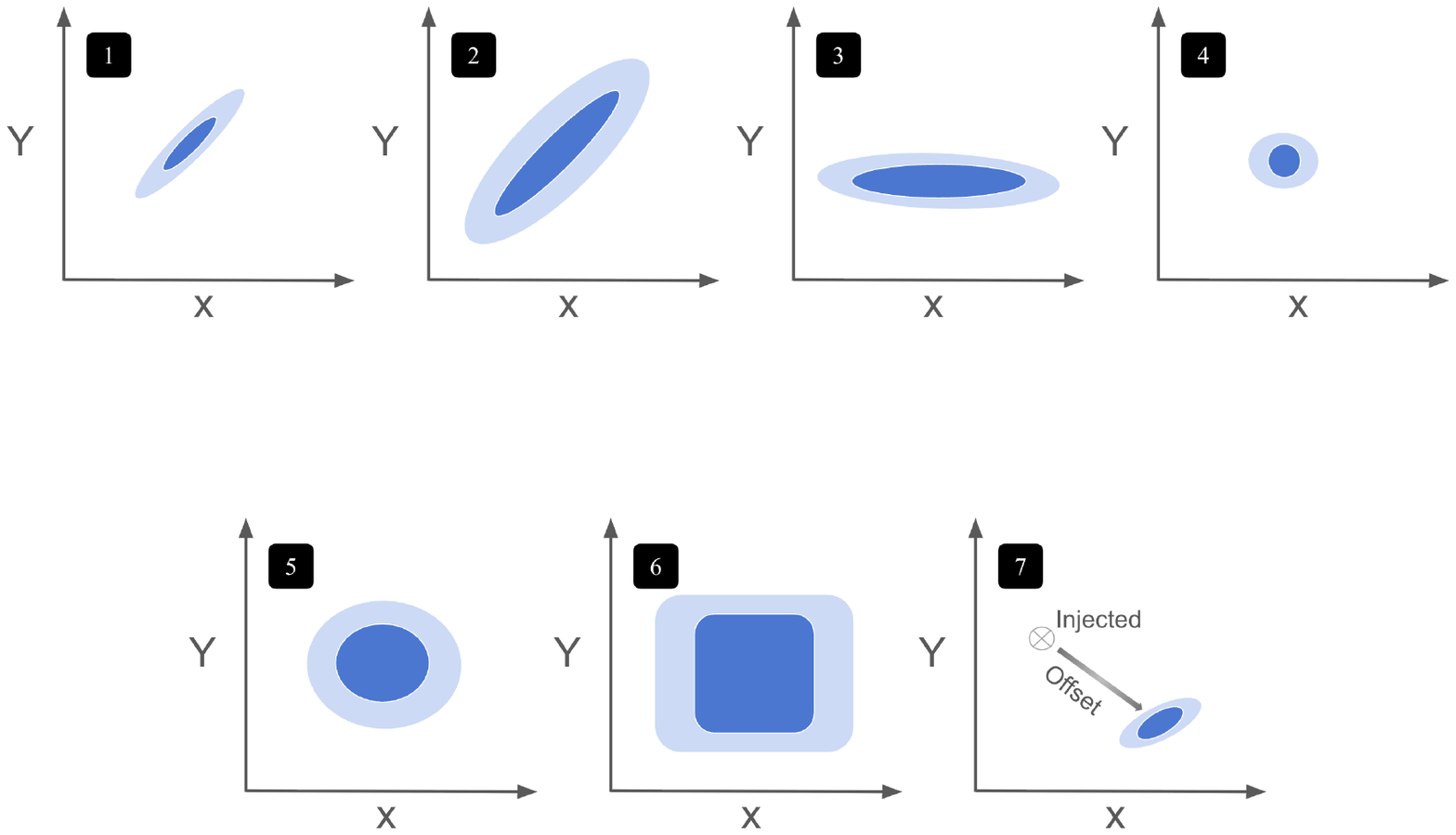}
    \caption{Schematic representation of posterior geometries outlined in Table \ref{tab:superstring_diagnostic_interpretation}.}
    \label{fig:pos_geo}
\end{figure*}
To quantify the reconstruction of cosmic-superstring parameters, we use the
posterior covariance matrix in the two-dimensional signal-parameter sector
\begin{equation}
    \theta
    =
    \left(
    \theta_1,\theta_2
    \right)
    =
    \left(
    \log_{10}G\mu,\log_{10}P
    \right).
\end{equation}
The same covariance diagnostics were developed and discussed in detail in
Ref.~\cite{Datta:2026hka}; here we summarize only the ingredients
needed to interpret the superstring reconstruction maps.

Let
\begin{equation}
    C_{ij}
    =
    {\rm Cov}(\theta_i,\theta_j)
\end{equation}
be the posterior covariance matrix in the sampled logarithmic coordinates. We
denote its larger and smaller eigenvalues by
\begin{equation}
    \lambda_\parallel=\lambda_+,
    \qquad
    \lambda_\perp=\lambda_- ,
\end{equation}
where \(\lambda_\parallel\) and \(\lambda_\perp\) correspond respectively to the
broad and narrow principal directions of the posterior. The posterior
anisotropy is measured by
\begin{equation}
    \kappa_{\rm deg}
    =
    \frac{\lambda_\parallel}{\lambda_\perp}.
\end{equation}
Large \(\kappa_{\rm deg}\) indicates that the data constrain one combination of
\((G\mu,P)\) much more accurately than the orthogonal combination. It does not,
by itself, imply poor reconstruction.

The total covariance area is characterized by
\begin{equation}
    A_{\rm cov}
    =
    \sqrt{\det C}
    =
    \sqrt{\lambda_\parallel\lambda_\perp}.
\end{equation}
This quantity measures posterior localization rather than posterior shape. A
posterior can therefore be highly elongated, \(\kappa_{\rm deg}\gg1\), and
still well localized if \(A_{\rm cov}\) is small. This is the case of compact
but correlated reconstruction.

We also compute the marginalized widths
\begin{equation}
    \sigma_{\log_{10}G\mu}=\sqrt{C_{11}},
    \qquad
    \sigma_{\log_{10}P}=\sqrt{C_{22}},
\end{equation}
and the Pearson correlation coefficient
\begin{equation}
    \rho
    =
    \frac{C_{12}}{\sqrt{C_{11}C_{22}}}.
\end{equation}
The coefficient \(\rho\) measures the orientation of the posterior relative to
the chosen coordinate axes, while \(\kappa_{\rm deg}\) measures the hierarchy of
principal variances. Thus \(|\rho|\simeq1\) indicates a correlated
\(G\mu\)--\(P\) trade-off, whereas \(\kappa_{\rm deg}\gg1\) indicates an
elongated posterior independently of its orientation.

For the amplitude-only superstring model, a large correlated elongation is
expected whenever the signal depends mainly on the effective amplitude
combination
\begin{equation}
    (G\mu)^q P^{-\beta}
    \simeq
    {\rm const.}
\end{equation}
In this regime, LISA measures one combination of \(G\mu\) and \(P\) accurately,
but cannot fully separate the string tension from the intercommutation
probability. In the cusp--kink mixture model, this degeneracy can be weakened
only if the change in harmonic structure produces measurable spectral-shape
information within the LISA band. \\Finally, reconstruction quality is assessed by combining covariance-based
precision diagnostics with accuracy diagnostics. The quantities
\(\kappa_{\rm deg}\), \(\rho\), \(\lambda_\perp\), \(A_{\rm cov}\), and the
marginalized widths characterize the posterior geometry and the degree of
localization in parameter space. On the other hand, the displacement between the
injected and recovered parameters measures reconstruction accuracy. A small
posterior area therefore does not by itself imply an unbiased recovery; it only
indicates that, within the assumed model, the data select a compact region of
parameter space. The diagnostic patterns used to interpret the reconstruction
maps are summarized in
Table~\ref{tab:superstring_diagnostic_interpretation}. A schematic representation of the posterior geometries outlined in Table~\ref{tab:superstring_diagnostic_interpretation} is presented in Fig. \ref{fig:pos_geo}.

\section{Resluts and discussion}
\label{sec:results_superstrings}

\subsection{Bayesian comparison of the two superstring models}
\label{subsec:superstring_bayes_factor}
\begin{figure}
    \centering
    \includegraphics[width=1\linewidth]{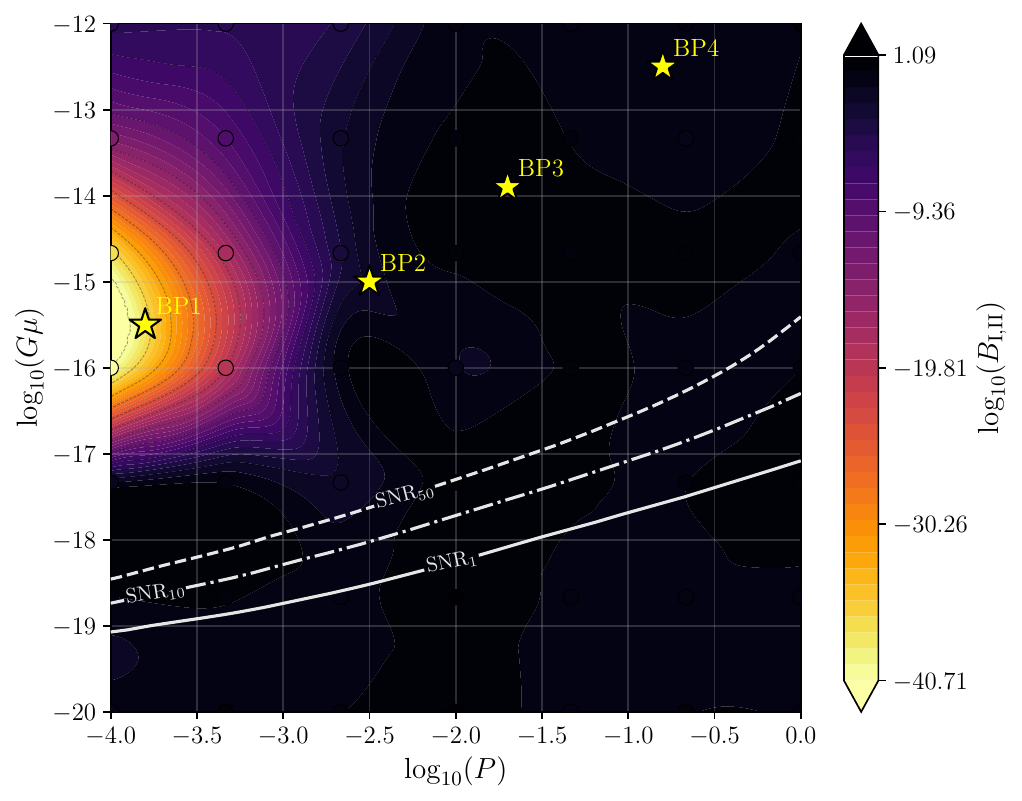}
    \caption{
Bayesian model comparison between the two cosmic-superstring signal models in
the \((G\mu,P)\) plane. The color scale shows
\(\log_{10}B_{1,2}=\log_{10}(Z_1/Z_2)\), where \(Z_1\) and \(Z_2\) are the
Bayesian evidences for Model I and Model II, respectively. The mock data are
generated using Model II. Negative values therefore indicate preference for the
shape-extended cusp--kink model, while values close to zero indicate that the
two models are not distinguishable by the LISA data. The overlaid contours show
the signal-to-noise ratio of the injected superstring background. At large
intercommutation probability, \(P\gg P_0\), Model II becomes cusp dominated and
approaches the amplitude-only Model-I spectrum, leading to
\(\log_{10}B_{1,2}\simeq0\). At small \(P\), the reduced intercommutation
probability enhances the signal amplitude, while the Model-II spectrum becomes
kink dominated, producing a strong preference for Model II whenever the
cusp--kink shape difference is measurable in the LISA band.
}
\label{fig:bf_superstrings}
\end{figure}
We begin by comparing the Bayesian evidences of the two superstring signal
models. For each injected point in the \((G\mu,P)\) plane, we generate mock LISA
data using Model II and analyze the same data with both Model I and Model II.
The model comparison is quantified by
\[
    \log_{10}B_{1,2}
    =
    \log_{10}\left(\frac{Z_1}{Z_2}\right),
\]
where \(Z_1\) and \(Z_2\) denote the evidences of the amplitude-only cusp model
and the cusp--kink mixture model, respectively. With this convention,
\(\log_{10}B_{1,2}<0\) indicates preference for the true injected model, Model
II, while \(\log_{10}B_{1,2}\simeq0\) indicates that the two models are
effectively indistinguishable.

Fig.~\ref{fig:bf_superstrings} shows that the Bayes factor is controlled not
only by the total signal-to-noise ratio, but also by the amount of spectral-shape
information available in the LISA band. At large intercommutation probability,
\(P\gg P_0\), the cusp fraction satisfies \(p_c(P)\simeq1\), and Model II
reduces to
\[
    \Omega_{\rm SS}^{\rm II}
    \simeq
    P^{-\beta}\Omega_{\rm cusp}.
\]
This is the same spectral form as Model I. Consequently, even when the signal is
detectable, the data do not strongly distinguish the two models and the Bayes
factor remains close to unity,
\[
    \log_{10}B_{1,2}\simeq0 .
\]

The situation changes at small \(P\). In this regime \(p_c(P)\ll1\), so the
injected Model-II signal becomes kink dominated,
\[
    \Omega_{\rm SS}^{\rm II}
    \simeq
    P^{-\beta}\Omega_{\rm kink},
\]
whereas Model I remains cusp dominated. The difference between the two spectra
is therefore approximately
\[
    \Delta\Omega_{\rm SS}(f)
    =
    \Omega_{\rm SS}^{\rm II}
    -
    \Omega_{\rm SS}^{\rm I}
    =
    P^{-\beta}
    \bigl[1-p_c(P)\bigr]
    \left[
        \Omega_{\rm kink}(f)-\Omega_{\rm cusp}(f)
    \right].
\]
Thus the strongest model-selection power appears where the reduced
intercommutation probability both enhances the overall amplitude and makes the
cusp--kink shape difference observable. This explains the strongly negative
region of \(\log_{10}B_{1,2}\): the data favor Model II because the
amplitude-only cusp model cannot reproduce the kink-dominated spectral shape.

This evidence-level comparison provides the first indication of where the
additional shape information in Model II is identifiable. We next examine the
posterior covariance diagnostics, including \(\kappa_{\rm deg}\) and
\(1-\rho^2\), to determine whether the parameters \((G\mu,P)\) are independently
localized or only constrained through correlated combinations within each model.
\subsection{Posterior reconstruction in Model I}
\label{subsec:model1_reconstruction}
\begin{figure*}
    \centering
    \includegraphics[width=0.48\linewidth]{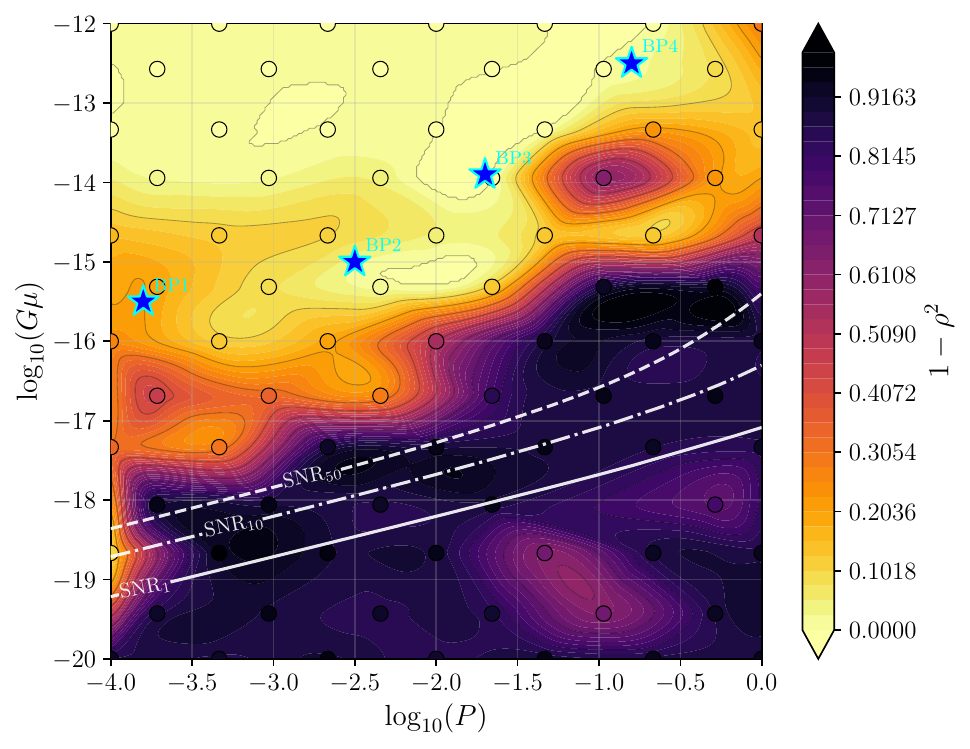} \includegraphics[width=0.48\linewidth]{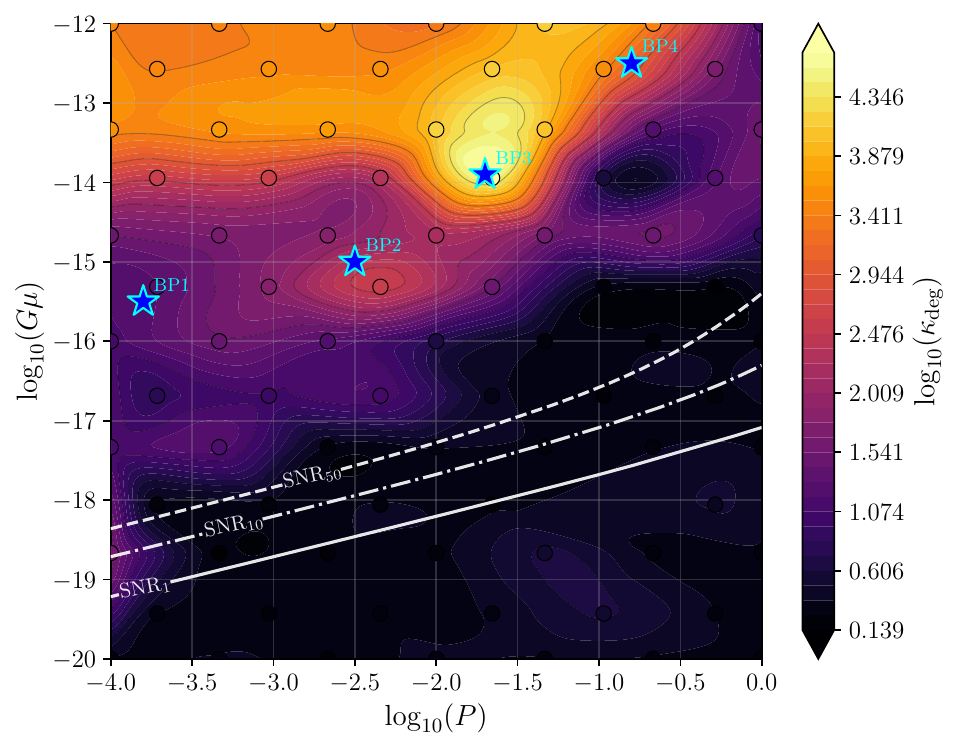}\\
    \includegraphics[width=0.48\linewidth]{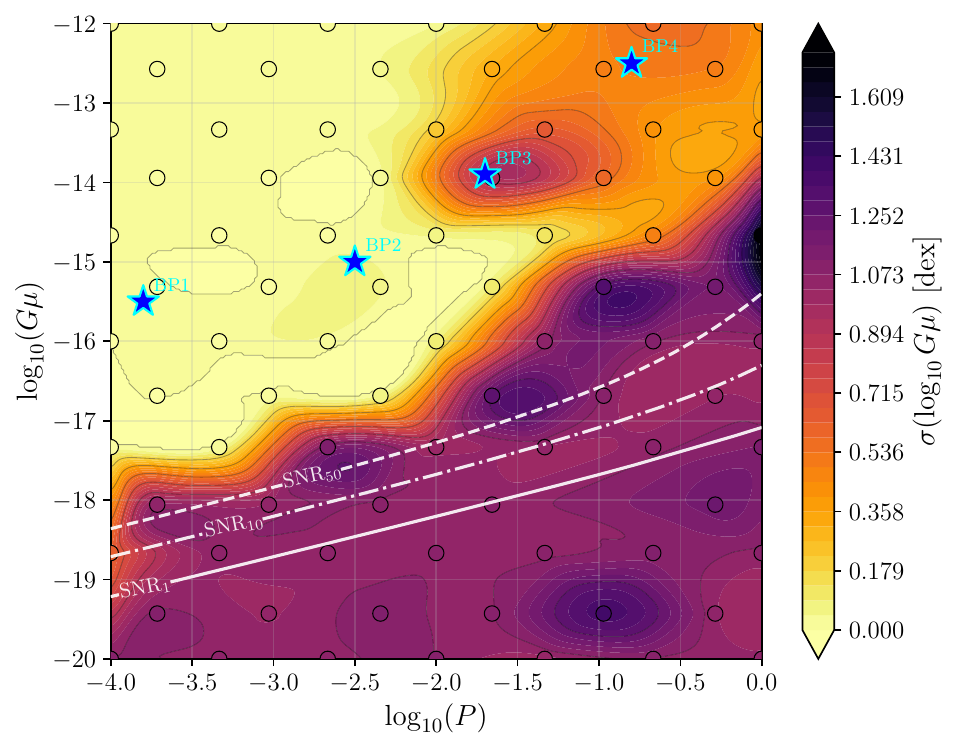}\includegraphics[width=0.48\linewidth]{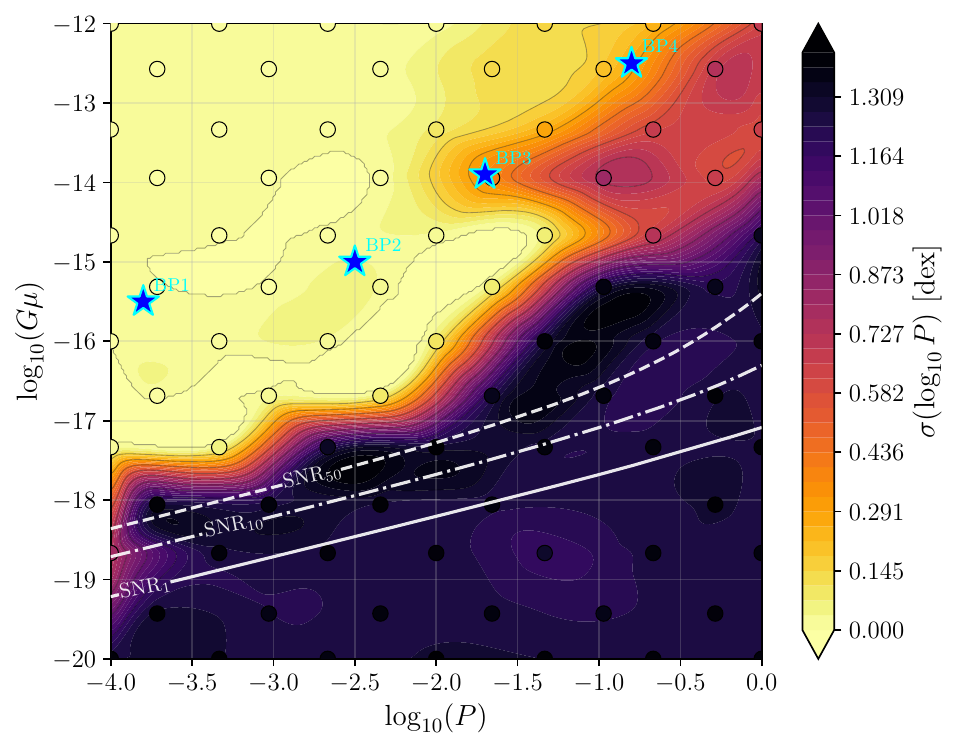}
  \caption{
Posterior-covariance diagnostics for Model I, the amplitude-only cosmic
superstring model, in the \((G\mu,P)\) plane. The panels show
\(1-\rho^2\), \(\kappa_{\rm deg}\), \(\sigma(\log_{10}G\mu)\), and
\(\sigma(\log_{10}P)\), where \(\rho\) is the correlation coefficient between
\(\log_{10}G\mu\) and \(\log_{10}P\), \(\kappa_{\rm deg}\) is the ratio of the
larger to smaller covariance eigenvalue, and the \(\sigma_i\) are marginalized
posterior widths in dex. The cyan stars mark the benchmark points BP1--BP4,
which are used later for the explicit posterior reconstructions, and the white
curves show the injected-signal SNR contours. Small \(1-\rho^2\) indicates a
highly correlated posterior, while large \(\kappa_{\rm deg}\) indicates strong
anisotropy in the covariance ellipse. The marginalized widths quantify whether
the individual parameters are localized, independently of the orientation of
the posterior.
}
\label{fig:model1_covariance_diagnostics}
\end{figure*}
\begin{figure*}
    \centering
    \includegraphics[width=0.5\linewidth]{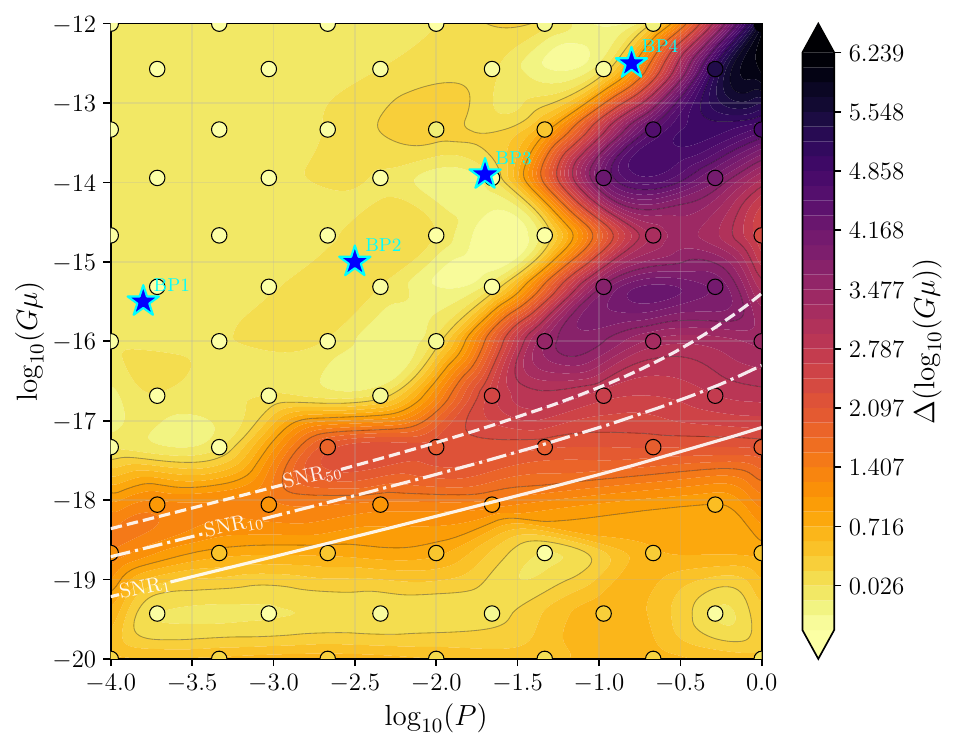}\includegraphics[width=0.5\linewidth]{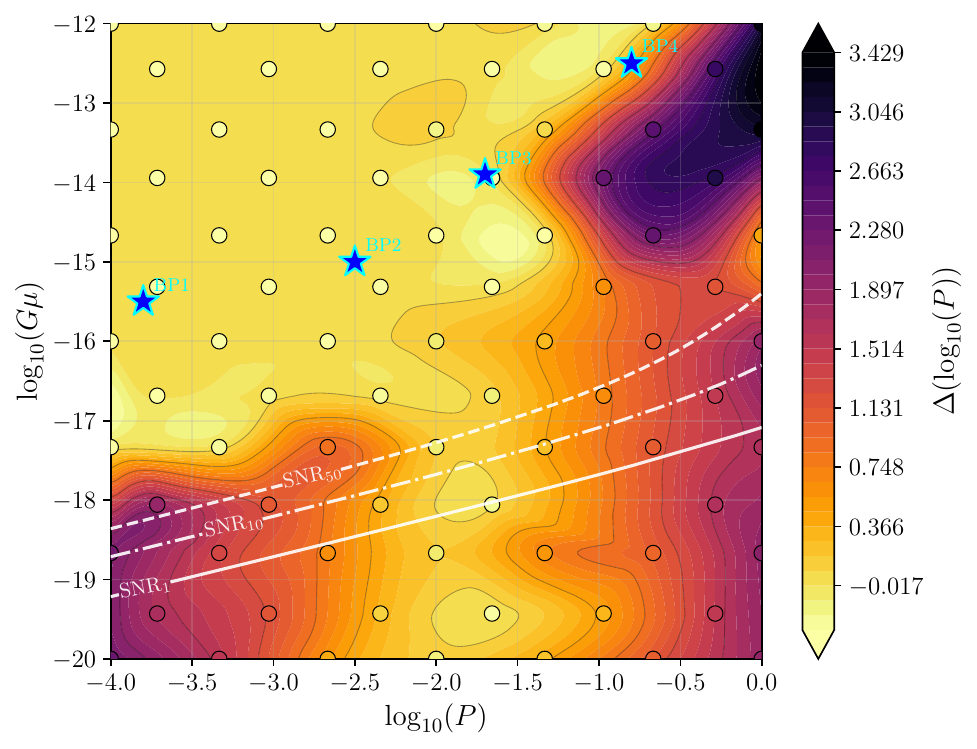}
   \caption{
Reconstruction bias for Model I in the \((G\mu,P)\) plane. The left panel shows
the displacement in \(\log_{10}G\mu\), while the right panel shows the
displacement in \(\log_{10}P\), between the injected and recovered posterior
locations. The cyan stars denote BP1--BP4 and the white curves show the SNR
contours. These maps test the accuracy of the reconstruction rather than its
precision: a point may have a small posterior covariance but still be biased if
the posterior is displaced from the injected value.
}
\label{fig:model1_bias_maps}
\end{figure*}
\begin{figure*}
    \centering
    \includegraphics[width=0.5\linewidth]{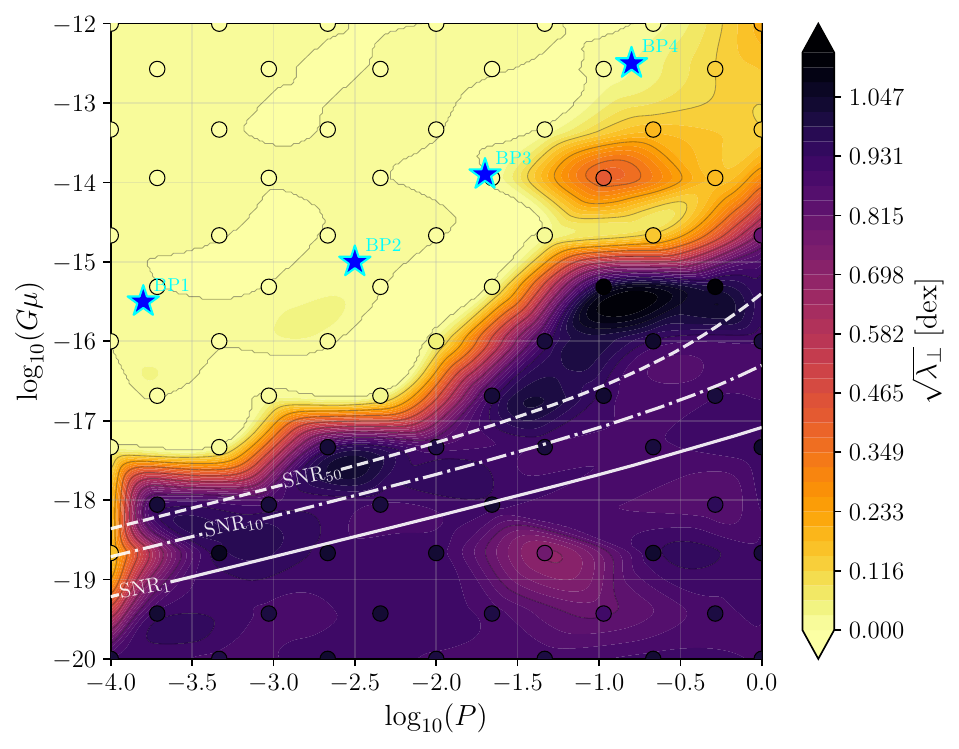}\includegraphics[width=0.5\linewidth]{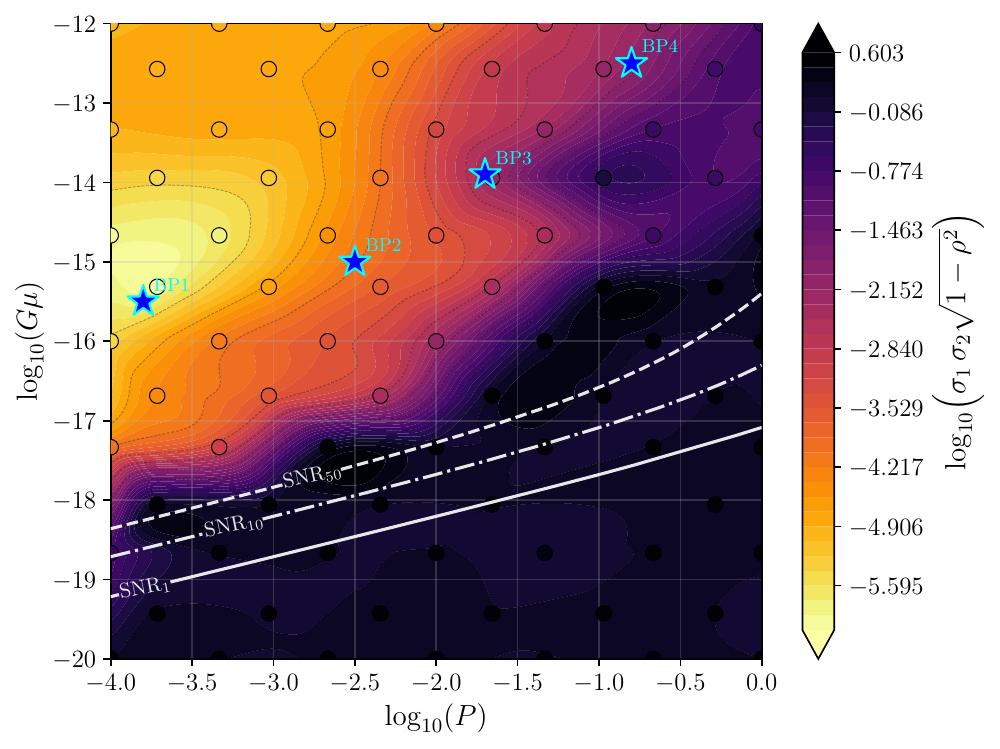}
    \caption{
Principal-width and covariance-area diagnostics for Model I. The left panel
shows \(\sqrt{\lambda_\perp}\), the posterior width along the best-constrained
principal direction in the \((\log_{10}G\mu,\log_{10}P)\) plane. The right
panel shows the covariance area
\(A_{\rm cov}=\sqrt{\det C}=\sigma_1\sigma_2\sqrt{1-\rho^2}\), which measures
the overall localization area of the two-dimensional posterior. The benchmark
points BP1--BP4 and the SNR contours are shown as in the previous figures.
Together these diagnostics distinguish compact correlated reconstruction from
broad degeneracy: \(\sqrt{\lambda_\perp}\) measures the best-constrained
parameter combination, while \(A_{\rm cov}\) measures the total posterior
localization.
}
\label{fig:model1_principal_area}
\end{figure*}
We first examine the reconstruction properties of Model I, in which the
intercommutation probability enters only through the amplitude rescaling
\[
    \Omega_{\rm SS}^{\rm I}
    =
    P^{-\beta}\Omega_{\rm cusp}(f;G\mu).
\]
Since \(P\) changes the overall normalization of the signal, while \(G\mu\)
changes both the amplitude and, to a lesser extent, the spectral position and
shape in the LISA band, one generically expects a correlated posterior in the
\((\log_{10}G\mu,\log_{10}P)\) plane. In the simplest local approximation,
where the spectrum scales as \(\Omega_{\rm cusp}\propto (G\mu)^q\), the data
primarily constrain the combination
\[
    q\log_{10}G\mu-\beta\log_{10}P \simeq {\rm const.}
\]
This produces the characteristic trade-off direction between \(G\mu\) and \(P\).

Fig.~\ref{fig:model1_covariance_diagnostics} shows this behavior explicitly.
Large regions of the detectable parameter space have small \(1-\rho^2\),
indicating strong correlation between \(\log_{10}G\mu\) and \(\log_{10}P\).
The corresponding \(\kappa_{\rm deg}\) map shows that the posterior is often
highly anisotropic: one linear combination of the two parameters is constrained
much more accurately than the orthogonal combination. This is the expected
signature of an amplitude degeneracy. Importantly, large \(\kappa_{\rm deg}\)
does not by itself imply a poor reconstruction. It only indicates that the
posterior ellipse is elongated. Whether the reconstruction is useful must be
judged together with the marginalized widths and the covariance area.

The maps of \(\sigma(\log_{10}G\mu)\) and \(\sigma(\log_{10}P)\) show where the
individual parameters are well localized. In the high-SNR region, especially
above the main detectability contours, both widths decrease, indicating that the
signal carries enough information to localize the amplitude combination. At the
same time, the persistence of strong correlation shows that the improvement is
not equivalent to fully independent measurements of \(G\mu\) and \(P\). Rather,
Model I often yields a compact but correlated reconstruction of the dominant
amplitude combination.

The bias maps in Fig.~\ref{fig:model1_bias_maps} provide a complementary test.
They show the displacement between the injected parameters and the recovered
posterior location. These maps should be interpreted separately from the
covariance diagnostics. A small posterior width measures precision under the
assumed model, whereas a small bias measures accuracy. Regions with large bias
therefore indicate where the posterior can be displaced even if the covariance
appears relatively small. Such behavior can arise from foreground correlations,
prior-boundary effects, or non-Gaussian posterior structure, and it motivates
the explicit benchmark posterior plots shown below.

\begin{figure*}
    \centering
\includegraphics[width=1\linewidth]{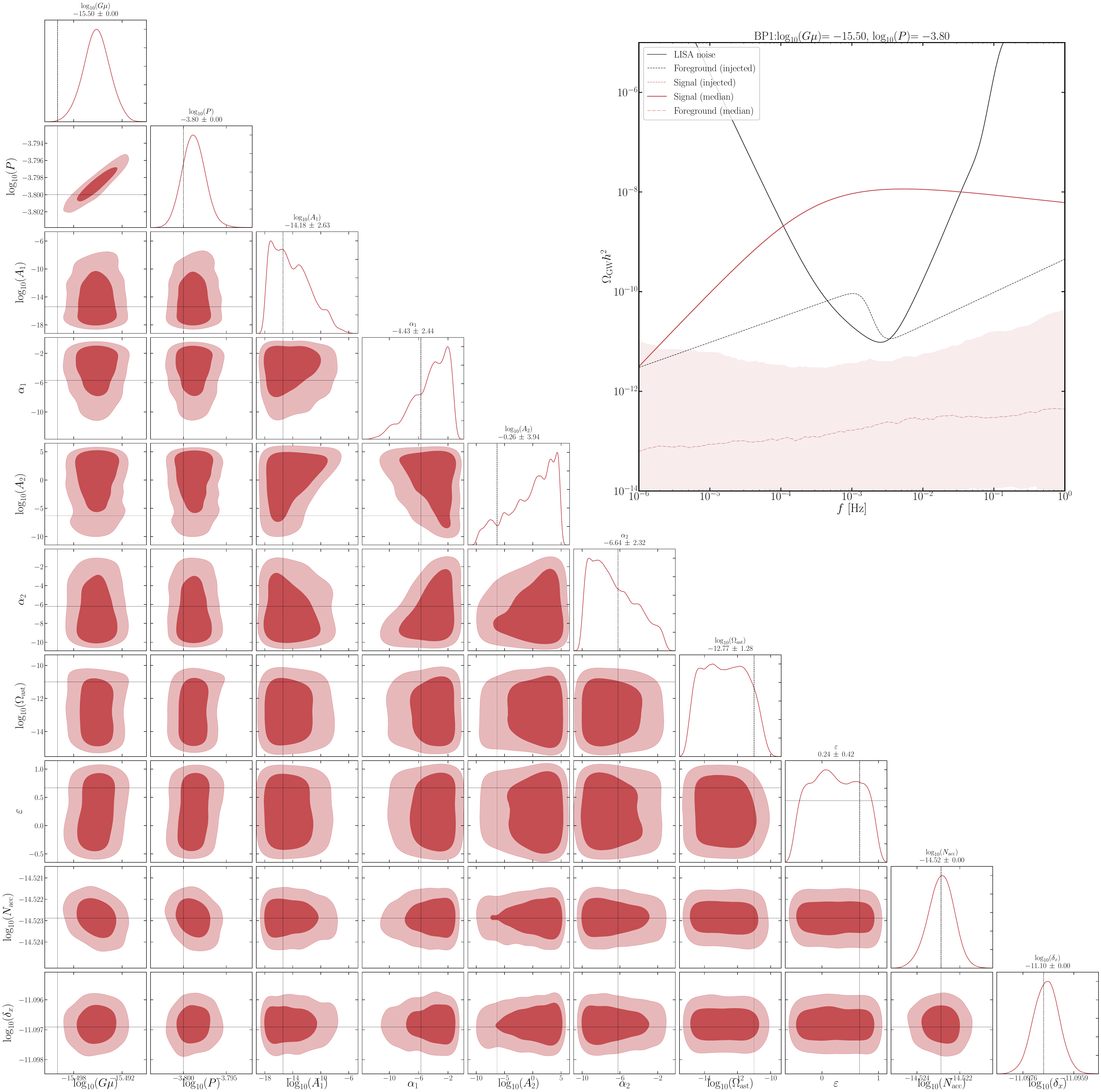}
   \caption{
Ten-dimensional corner plot for BP1 in Model I, corresponding to the
amplitude-only cusp-dominated cosmic-superstring scenario. The posterior is
shown for the full parameter set, including the signal parameters
\((G\mu,P)\), instrumental-noise parameters, unresolved extragalactic
foreground parameters, and the four parameters of the flexible Galactic
double-white-dwarf foreground template. The flexible Galactic foreground allows
both amplitude and spectral-shape freedom, providing a conservative test of
cosmic-superstring reconstruction.
}
\label{fig:corner_bp1_model1}
\end{figure*}

\begin{figure*}
    \centering \includegraphics[width=1\linewidth]{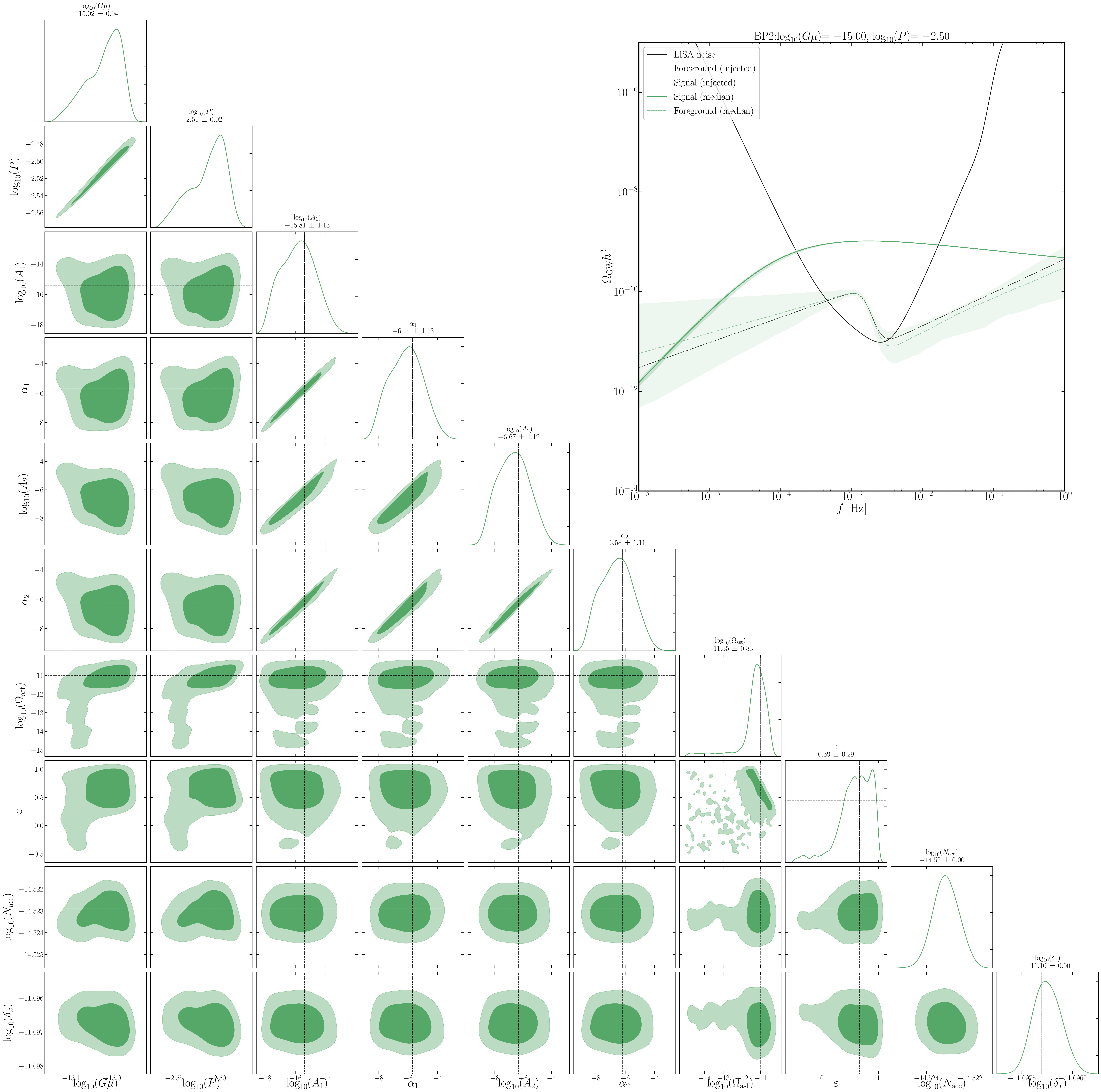}
   \caption{
Ten-dimensional corner plot for BP2 in Model I, corresponding to the
amplitude-only cusp-dominated cosmic-superstring scenario. The posterior is
shown for the full parameter set, including the signal parameters
\((G\mu,P)\), instrumental-noise parameters, unresolved extragalactic
foreground parameters, and the four parameters of the flexible Galactic
double-white-dwarf foreground template. The flexible Galactic foreground allows
both amplitude and spectral-shape freedom, providing a conservative test of
cosmic-superstring reconstruction.
}
\label{fig:corner_bp2_model1}
\end{figure*}

\begin{figure*}
    \centering
\includegraphics[width=1\linewidth]{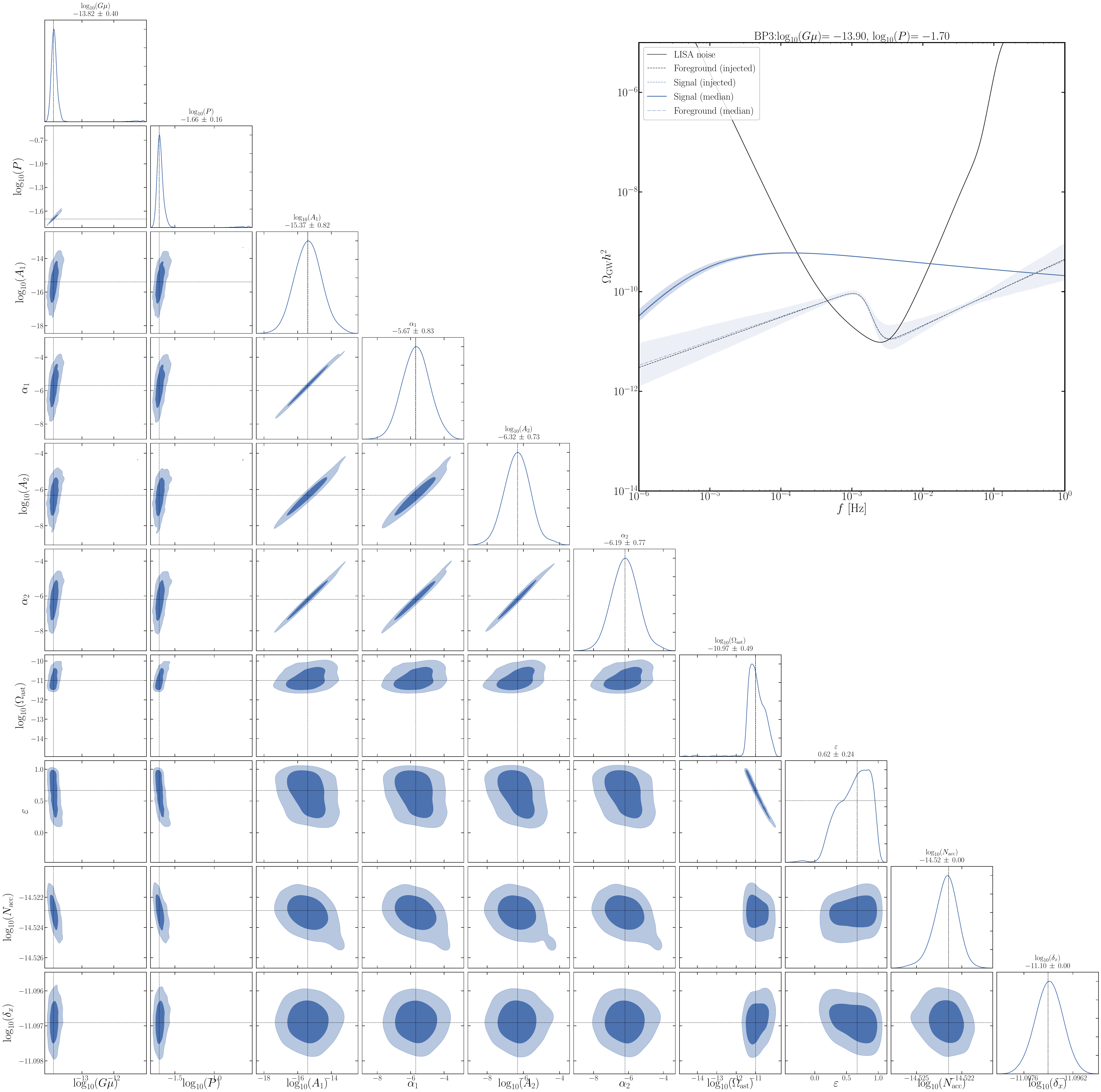}
   \caption{
Ten-dimensional corner plot for BP3 in Model I, corresponding to the
amplitude-only cusp-dominated cosmic-superstring scenario. The posterior is
shown for the full parameter set, including the signal parameters
\((G\mu,P)\), instrumental-noise parameters, unresolved extragalactic
foreground parameters, and the four parameters of the flexible Galactic
double-white-dwarf foreground template. The flexible Galactic foreground allows
both amplitude and spectral-shape freedom, providing a conservative test of
cosmic-superstring reconstruction.
}
\label{fig:corner_bp3_model1}
\end{figure*}

\begin{figure*}
    \centering
\includegraphics[width=1\linewidth]{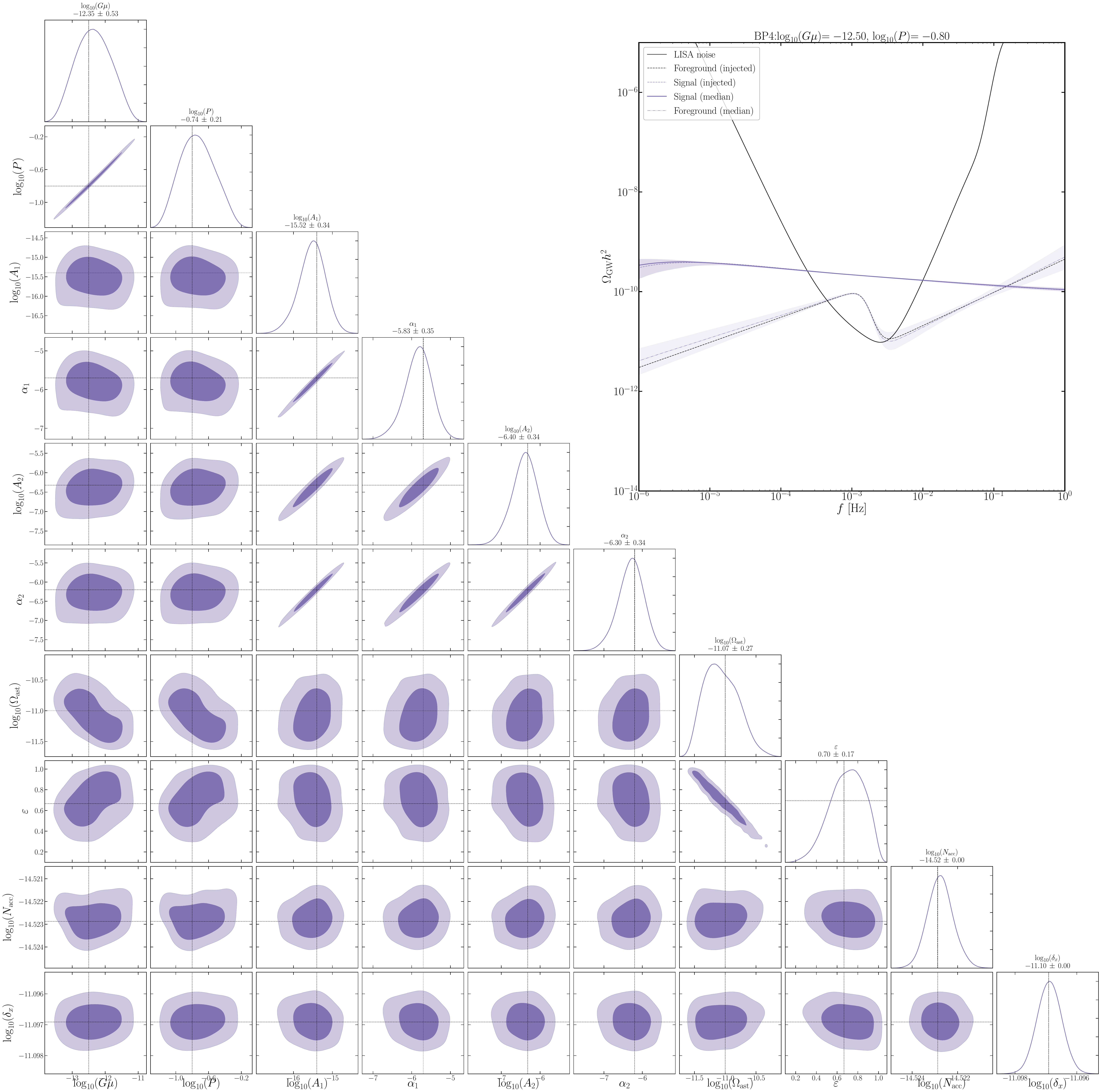}
  \caption{
Ten-dimensional corner plot for BP4 in Model I, corresponding to the
amplitude-only cusp-dominated cosmic-superstring scenario. The posterior is
shown for the full parameter set, including the signal parameters
\((G\mu,P)\), instrumental-noise parameters, unresolved extragalactic
foreground parameters, and the four parameters of the flexible Galactic
double-white-dwarf foreground template. The flexible Galactic foreground allows
both amplitude and spectral-shape freedom, providing a conservative test of
cosmic-superstring reconstruction.
}
\label{fig:corner_bp4_model1}
\end{figure*}

Finally, Fig.~\ref{fig:model1_principal_area} separates the best-constrained
principal direction from the total two-dimensional localization. The map of
\(\sqrt{\lambda_\perp}\) identifies where LISA tightly measures at least one
combination of \((G\mu,P)\), while \(A_{\rm cov}\) quantifies the full posterior
area. A small \(\sqrt{\lambda_\perp}\) together with a large
\(\kappa_{\rm deg}\) corresponds to a thin correlated posterior ridge. If
\(A_{\rm cov}\) is also small, this is a compact correlated reconstruction; if
\(A_{\rm cov}\) remains large, the result is instead a broad degeneracy. The
benchmark points BP1--BP4 are chosen to illustrate these different regimes in
the explicit posterior distributions in Fig.s \ref{fig:corner_bp1_model1}-\ref{fig:corner_bp4_model1}.

\subsection{Posterior reconstruction in Model II}
\label{subsec:model2_reconstruction}
\begin{figure*}
    \centering
    \includegraphics[width=0.48\linewidth]{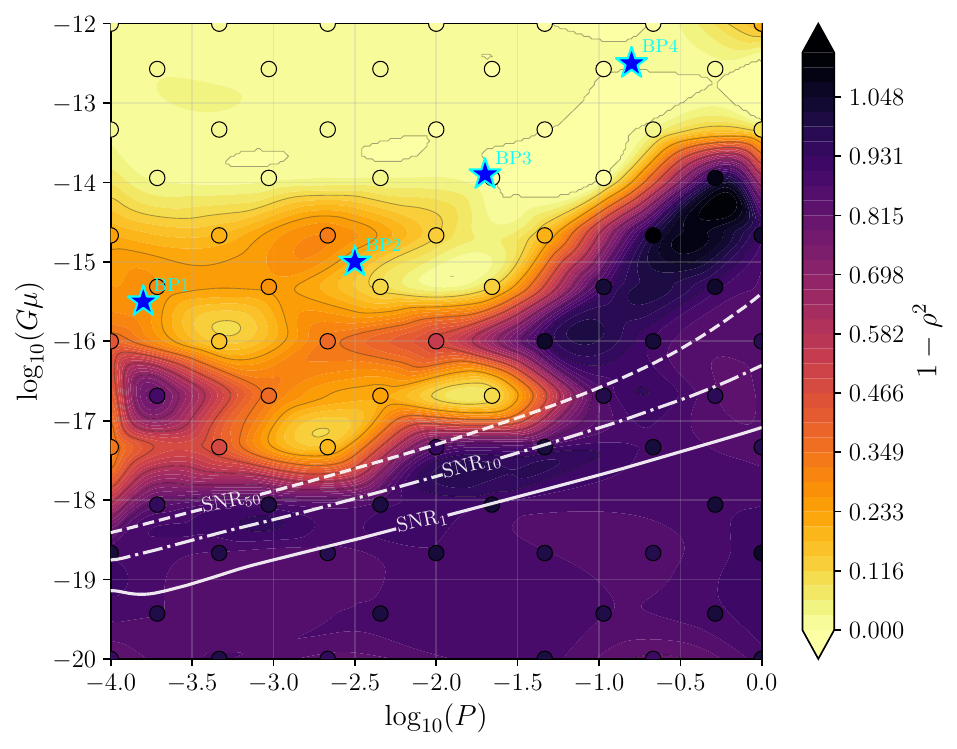} \includegraphics[width=0.48\linewidth]{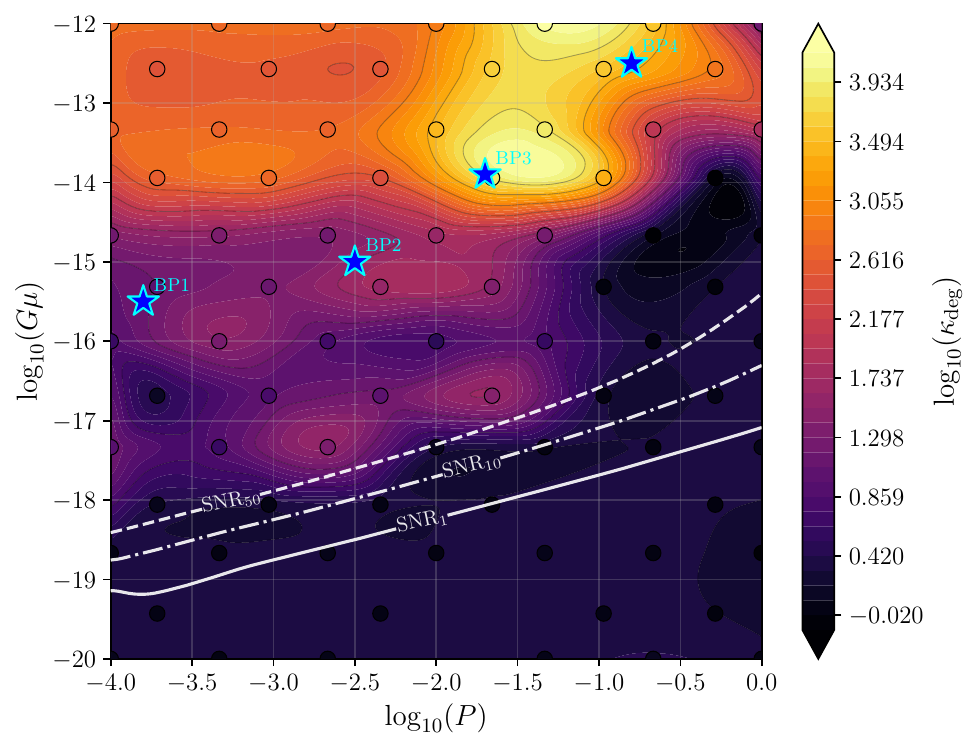}\\
    \includegraphics[width=0.48\linewidth]{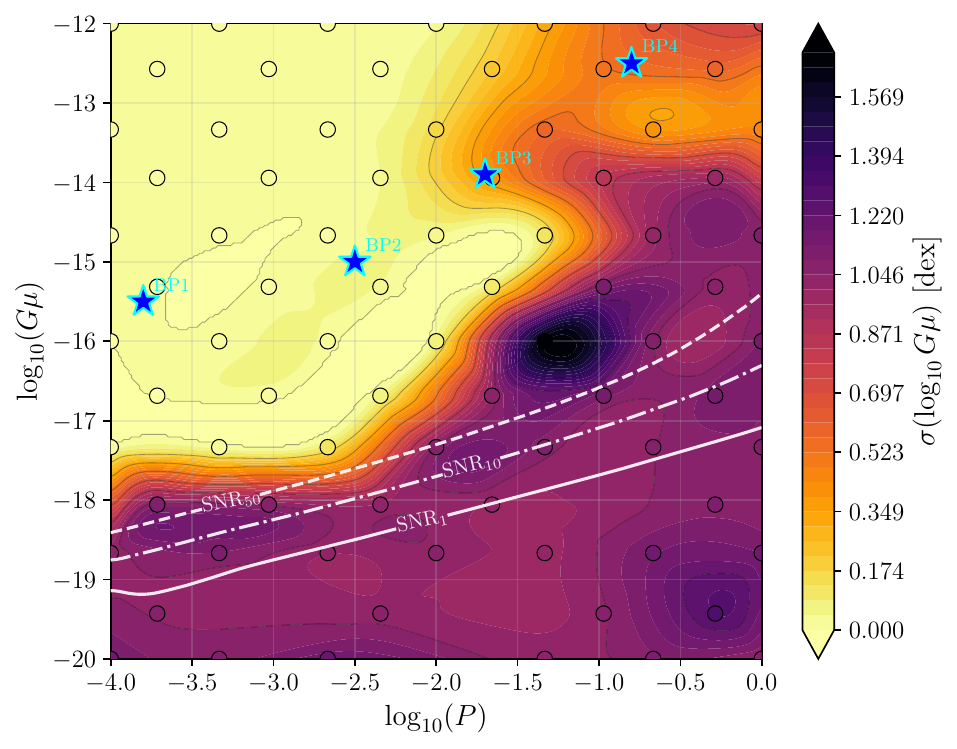}\includegraphics[width=0.48\linewidth]{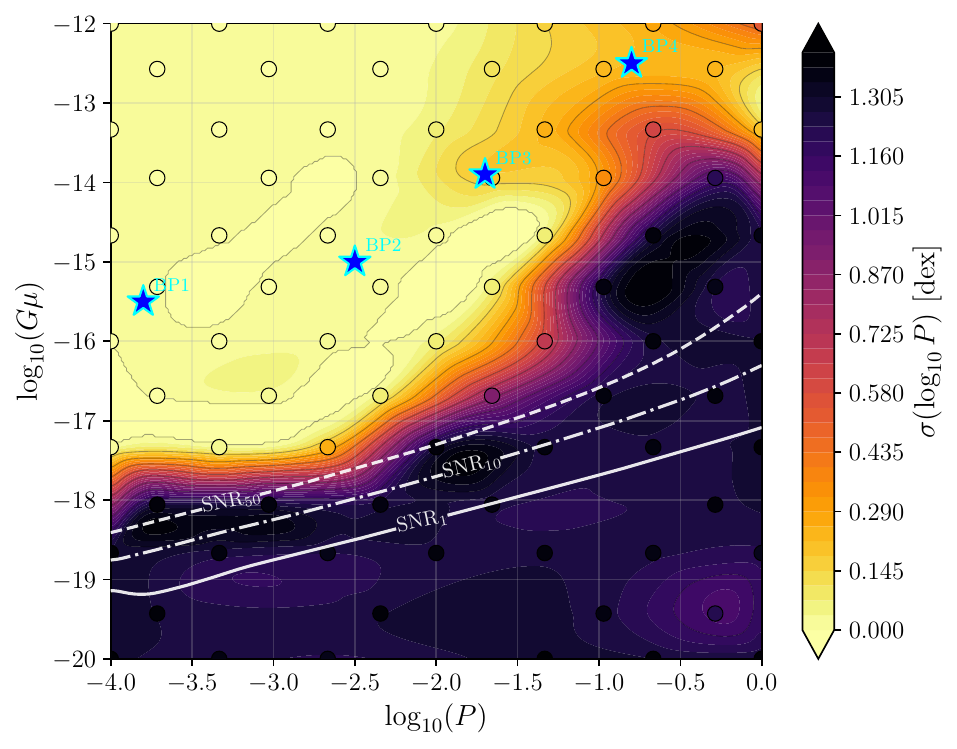}
  \caption{
Posterior-covariance diagnostics for Model II, the amplitude-only cosmic
superstring model, in the \((G\mu,P)\) plane. The panels show
\(1-\rho^2\), \(\kappa_{\rm deg}\), \(\sigma(\log_{10}G\mu)\), and
\(\sigma(\log_{10}P)\), where \(\rho\) is the correlation coefficient between
\(\log_{10}G\mu\) and \(\log_{10}P\), \(\kappa_{\rm deg}\) is the ratio of the
larger to smaller covariance eigenvalue, and the \(\sigma_i\) are marginalized
posterior widths in dex. The cyan stars mark the benchmark points BP1--BP4,
which are used later for the explicit posterior reconstructions, and the white
curves show the injected-signal SNR contours. Small \(1-\rho^2\) indicates a
highly correlated posterior, while large \(\kappa_{\rm deg}\) indicates strong
anisotropy in the covariance ellipse. The marginalized widths quantify whether
the individual parameters are localized, independently of the orientation of
the posterior.
}
\label{fig:model2_covariance_diagnostics}
\end{figure*}
\begin{figure*}
    \centering
    \includegraphics[width=0.5\linewidth]{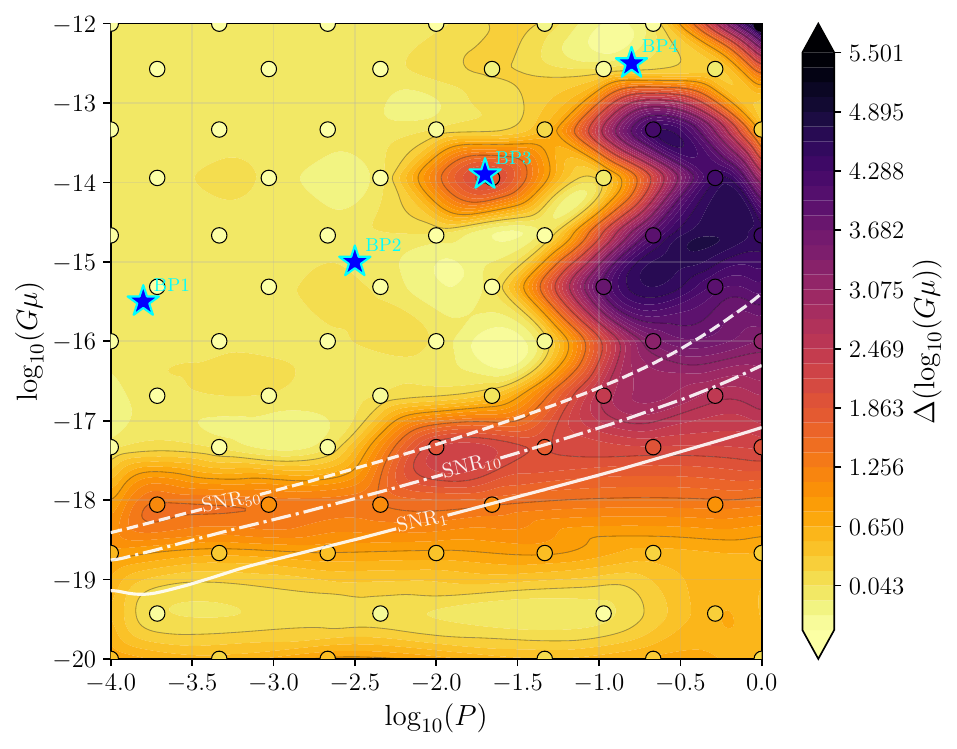}\includegraphics[width=0.5\linewidth]{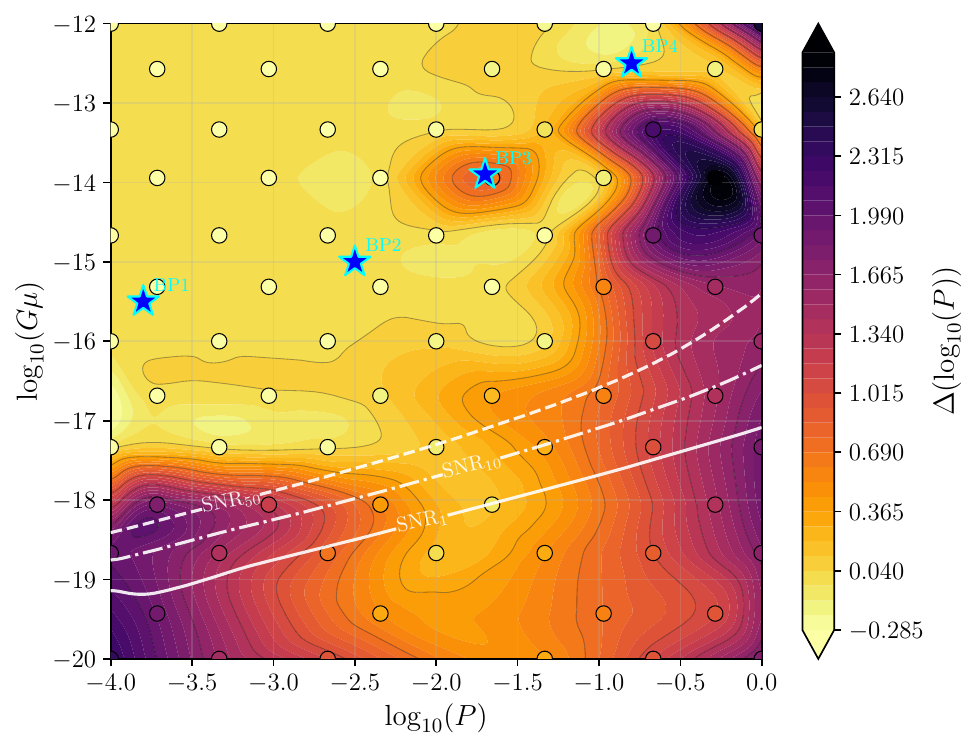}
   \caption{
Reconstruction bias for Model II in the \((G\mu,P)\) plane. The left panel shows
the displacement in \(\log_{10}G\mu\), while the right panel shows the
displacement in \(\log_{10}P\), between the injected and recovered posterior
locations. The cyan stars denote BP1--BP4 and the white curves show the SNR
contours. These maps test the accuracy of the reconstruction rather than its
precision: a point may have a small posterior covariance but still be biased if
the posterior is displaced from the injected value.
}
\label{fig:model2_bias_maps}
\end{figure*}
\begin{figure*}
    \centering
    \includegraphics[width=0.5\linewidth]{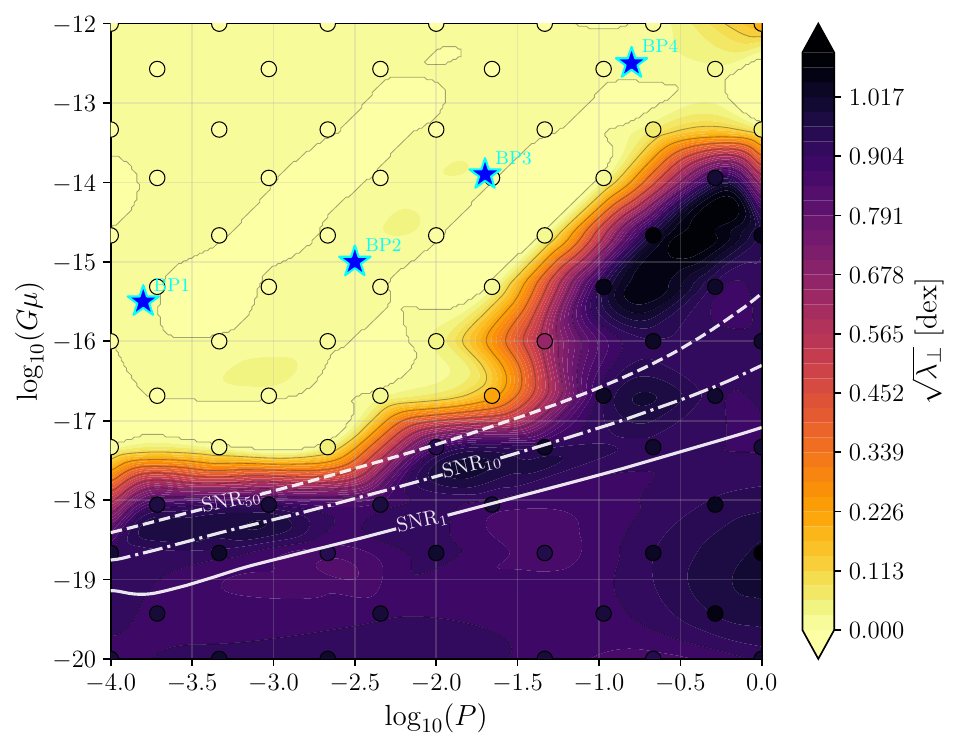}\includegraphics[width=0.5\linewidth]{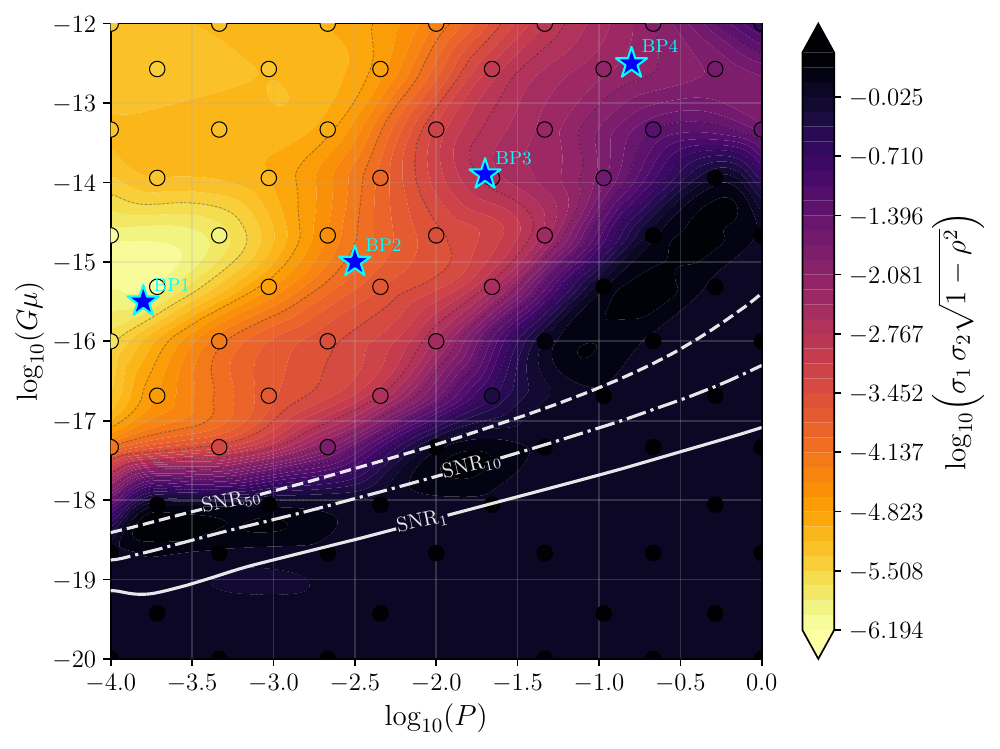}
    \caption{
Principal-width and covariance-area diagnostics for Model II. The left panel
shows \(\sqrt{\lambda_\perp}\), the posterior width along the best-constrained
principal direction in the \((\log_{10}G\mu,\log_{10}P)\) plane. The right
panel shows the covariance area
\(A_{\rm cov}=\sqrt{\det C}=\sigma_1\sigma_2\sqrt{1-\rho^2}\), which measures
the overall localization area of the two-dimensional posterior. The benchmark
points BP1--BP4 and the SNR contours are shown as in the previous figures.
Together these diagnostics distinguish compact correlated reconstruction from
broad degeneracy: \(\sqrt{\lambda_\perp}\) measures the best-constrained
parameter combination, while \(A_{\rm cov}\) measures the total posterior
localization.
}
\label{fig:model2_principal_area}
\end{figure*}
We now turn to Model II, in which the intercommutation probability affects both
the overall amplitude and the spectral shape through the cusp--kink mixture,
\[
    \Omega_{\rm SS}^{\rm II}
    =
    P^{-\beta}
    \left[
        p_c\,\Omega_{\rm cusp}
        +(1-p_c)\Omega_{\rm kink}
    \right].
\]
This additional shape dependence changes the interpretation of the
\((G\mu,P)\) posterior. In Model I, \(P\) acts only as an amplitude parameter,
so the dominant degeneracy follows an approximately constant-amplitude
direction. In Model II, the same amplitude degeneracy is still present, but it
can be weakened wherever changes in \(P\) also produce measurable changes in the
relative cusp and kink contributions.

Fig.~\ref{fig:model2_covariance_diagnostics} shows that the posterior remains
strongly correlated over a large part of the detectable parameter space. This is
expected: even in the cusp--kink model, the factor \(P^{-\beta}\) produces a
large amplitude response, and the data can still constrain an amplitude-like
combination of \(G\mu\) and \(P\). However, the regions where the marginalized
widths decrease and where \(1-\rho^2\) increases indicate that the spectral
shape contribution of \(P\) is becoming informative. In these regions, the
posterior is not determined solely by the normalization of the stochastic
background, but also by the frequency dependence induced by the cusp--kink
mixture.

The \(\kappa_{\rm deg}\) map should be interpreted together with the width maps.
Large \(\kappa_{\rm deg}\) indicates that the posterior ellipse is elongated,
but not necessarily that the reconstruction is poor. A large anisotropy combined
with small marginalized widths or small covariance area corresponds to a compact
but correlated reconstruction. By contrast, large anisotropy together with large
widths signals a broad degeneracy. Thus, as in Model I, the key distinction is
between measuring one parameter combination precisely and independently
localizing both \(G\mu\) and \(P\).

The bias maps in Fig.~\ref{fig:model2_bias_maps} show where the recovered
posterior location is displaced from the injection. These maps are especially
important for Model II because the signal model contains genuine spectral-shape
information. A small covariance region does not by itself guarantee that the
posterior is centered on the true point. Large displacements can occur when the
foreground model absorbs part of the cusp--kink spectral difference, when the
posterior is non-Gaussian, or when the recovered point is affected by prior
boundaries. 

Finally, Fig.~\ref{fig:model2_principal_area} separates the best-constrained
direction from the total two-dimensional localization. The map of
\(\sqrt{\lambda_\perp}\) identifies where LISA measures at least one
combination of \(G\mu\) and \(P\) accurately, while the covariance area measures
whether the full posterior is compact. In the regions where Model II was favored
by the Bayes-factor comparison, these diagnostics indicate whether that model
preference is accompanied by improved parameter reconstruction, or whether LISA
detects the shape-extended model while still leaving a correlated
\(G\mu\)--\(P\) degeneracy.
\section{Discussion and conclusions}
\label{sec:conclusions}

In this work we have studied the reconstruction of cosmic-superstring stochastic
gravitational-wave backgrounds with LISA. The analysis was designed to address
two related questions. First, can LISA distinguish a purely amplitude-rescaled
superstring spectrum from a model in which the intercommutation probability also
changes the spectral shape? Second, once a signal model is assumed, can the
underlying parameters $(G\mu,P)$ be reconstructed independently, or are they
constrained only through a correlated combination?

We considered two phenomenological descriptions. In Model I, reduced
intercommutation changes only the amplitude of a cusp-dominated spectrum,
\begin{equation}
    \Omega_{\rm SS}^{\rm I}
    =
    P^{-\beta}\Omega_{\rm cusp}.
\end{equation}
In Model II, the signal contains an additional cusp--kink mixture,
\begin{equation}
    \Omega_{\rm SS}^{\rm II}
    =
    P^{-\beta}
    \left[
        p_c\,\Omega_{\rm cusp}
        +(1-p_c)\Omega_{\rm kink}
    \right],
\end{equation}
with $p_c=p_c(P)$. Thus, in Model II, $P$ affects both the normalization and
the frequency dependence of the stochastic background. This distinction is
central to the inference problem: Model I tests whether LISA can measure the
amplitude effect of reduced intercommutation, while Model II tests whether LISA
can access additional spectral-shape information associated with the small-scale
structure of loops.

The Bayesian model-comparison results show that the two descriptions are not
distinguishable everywhere in the $(G\mu,P)$ plane. In the conservative
foreground analysis, the Bayes-factor map shows that the two models become hard
to separate when the observable spectrum is effectively cusp dominated. At large
intercommutation probability, $P\gg P_0$, the cusp fraction approaches unity,
$p_c(P)\simeq 1$, and Model II reduces to the same cusp-dominated form as
Model I. In this regime the Bayes factor remains close to unity even when the
signal itself is detectable. This is an important point: detectability of a
superstring background does not automatically imply distinguishability of the
underlying superstring phenomenology. If the observable spectrum is effectively
cusp dominated, LISA measures the presence of a stochastic signal but has little
leverage to decide whether the more general cusp--kink model is required.

The situation changes at smaller $P$. In this regime $p_c(P)\ll 1$, so the
Model-II signal becomes increasingly kink dominated, while Model I remains cusp
dominated. The spectral difference between the two models is controlled by
\begin{equation}
    \Delta\Omega_{\rm SS}(f)
    =
    P^{-\beta}
    \bigl[1-p_c(P)\bigr]
    \left[
        \Omega_{\rm kink}(f)-\Omega_{\rm cusp}(f)
    \right].
\end{equation}
Model discrimination is therefore strongest where three conditions are met: the
reduced intercommutation probability enhances the signal amplitude, the
cusp--kink mixture is sufficiently far from the pure-cusp limit, and the
resulting shape difference lies in the LISA-sensitive part of the spectrum. In
these regions the evidence can strongly favor Model II over the amplitude-only
description. The Bayes-factor map therefore traces detectable shape mismatch,
not simply total signal-to-noise ratio.

The posterior-covariance diagnostics clarify what happens after a model is
selected or assumed. In Model I, the dominant effect of $P$ is an amplitude
rescaling. Locally, if $\Omega_{\rm cusp}\propto (G\mu)^q$, the data constrain
approximately
\begin{equation}
    q\log_{10}G\mu-\beta\log_{10}P
    \simeq
    {\rm const.}
\end{equation}
This produces the expected correlated posterior geometry in the
$(\log_{10}G\mu,\log_{10}P)$ plane. Large values of $\kappa_{\rm deg}$ and
small values of $1-\rho^2$ identify regions where the posterior is elongated
along this trade-off direction. Such elongation does not by itself imply a poor
measurement. If the covariance area is small, the result is a compact but
correlated reconstruction: LISA measures one combination of $G\mu$ and $P$
accurately, even if the individual parameters remain partially degenerate.

Model II modifies this picture but does not remove the degeneracy everywhere.
The factor $P^{-\beta}$ still produces a strong amplitude response, so an
amplitude-like $G\mu$--$P$ correlation remains over much of the parameter space.
However, in regions where the cusp--kink mixture changes the spectral
morphology inside the LISA band, the posterior widths decrease and the
correlation can be weakened. In these regions, $P$ is no longer measured only
through the overall normalization; it also leaves a shape imprint. This is the
most favorable case for genuine two-parameter reconstruction of $(G\mu,P)$.

The bias maps provide an additional and necessary diagnostic. A small posterior
area measures precision under the assumed model, but it does not guarantee that
the posterior is centered on the injected parameters. Regions with sizable
biases indicate that the reconstruction can be accurate in some parts of the
plane and shifted in others. Such shifts may arise from residual degeneracies
with astrophysical foregrounds, instrumental-noise parameters, non-Gaussian
posterior structure, or prior-boundary effects. For this reason, the benchmark
posterior plots are essential: they show explicitly whether a point corresponds
to a broad degeneracy, a compact correlated reconstruction, or a precise but
biased recovery.

\begin{figure}
    \centering
    \includegraphics[width=1\linewidth]{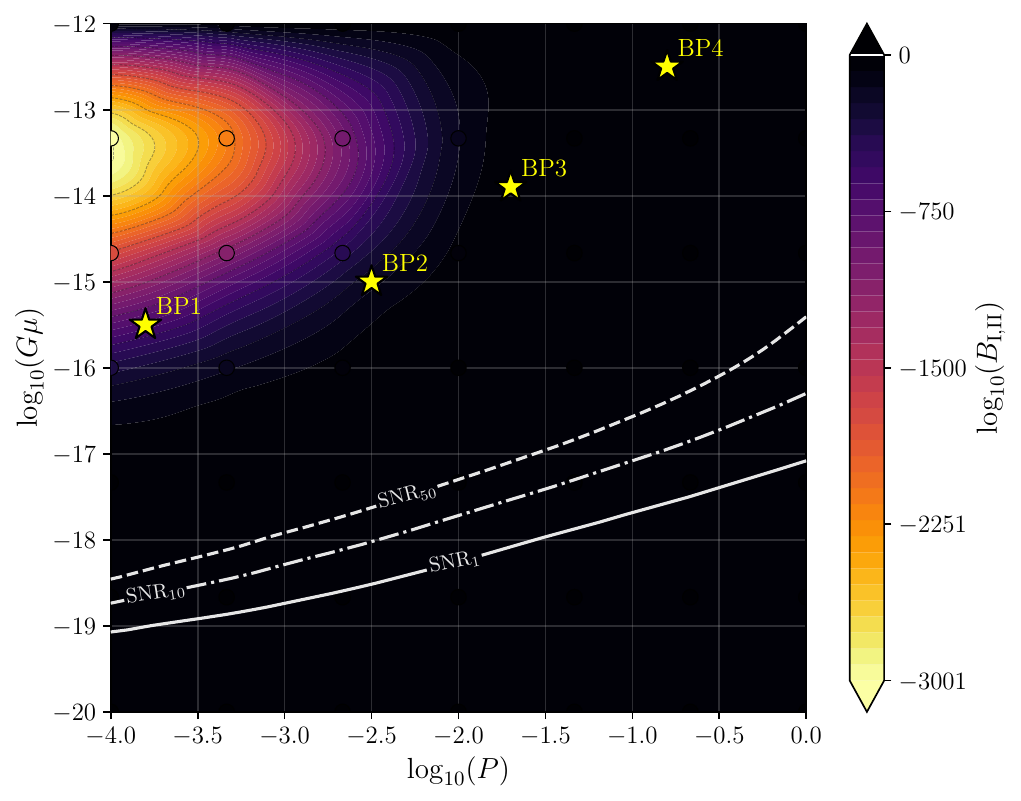}
   \caption{
Bayesian model comparison between the amplitude-only cosmic-superstring model
and the cusp--kink mixture model using the reduced tanh Galactic foreground
template. The color scale shows
\(\log_{10}B_{\rm I,II}=\log_{10}(Z_{\rm I}/Z_{\rm II})\), where \(Z_{\rm I}\)
and \(Z_{\rm II}\) are the Bayesian evidences for Model I and Model II,
respectively. The mock data are generated with Model II, so negative values
indicate preference for the shape-extended cusp--kink model. Compared with the
conservative Galactic foreground template, the reduced tanh description gives
the foreground less freedom to absorb broad spectral curvature, making the
cusp--kink shape difference more visible to LISA. As a result, the region in
which Model II is favored extends toward larger intercommutation probabilities
\(P\). The white curves show the injected-signal SNR contours.
}
\label{fig:bf_map_tanh}
\end{figure}

Throughout the main analysis we have used the conservative Galactic foreground
description, in which the unresolved double-white-dwarf component is allowed
both amplitude and spectral-shape freedom. This choice deliberately avoids
overestimating the reconstruction power of LISA. To illustrate the impact of
foreground assumptions on model discrimination, we also computed the Bayes-factor
map for the reduced tanh Galactic template shown in
Fig.~\ref{fig:bf_map_tanh}. We do not repeat the full set of covariance,
uncertainty, and bias maps for that case here; nevertheless, the Bayes-factor
result already makes the trend clear. Relative to the conservative foreground
model, the tanh-template case improves the distinguishability between Model I
and Model II and extends the informative region toward larger values of $P$.

The reason is physically transparent. In the conservative analysis, the Galactic
foreground has sufficient spectral freedom to absorb part of the broad curvature
associated with the cusp--kink mixture, especially when the deviation from the
pure-cusp limit is modest. In the reduced tanh model, by contrast, the
foreground shape is effectively fixed, and only its overall normalization is
varied. The foreground can then mimic the cosmological signal much less
efficiently. As a result, spectral differences between
$\Omega_{\rm cusp}$ and $\Omega_{\rm kink}$ survive more clearly in the
likelihood, so the evidence in favor of the shape-extended superstring model
increases. This is precisely why the Bayes-factor map in the tanh case begins to
accommodate larger $P$ values: even partially cusp-dominated spectra can remain
distinguishable once the Galactic foreground is prevented from absorbing the
relevant shape information. The tanh-template result should therefore be viewed
as an optimistic counterpart to our conservative baseline forecast.

Our main conclusions are as follows. First, LISA can distinguish the two
superstring signal models only where the cusp--kink spectral difference is both
present and measurable. At large $P$, Model II approaches the pure-cusp limit
and becomes effectively indistinguishable from Model I. Second, reduced
intercommutation generically induces strong correlations between $G\mu$ and $P$,
because both parameters affect the amplitude of the stochastic background.
Third, the additional shape dependence in Model II can weaken this degeneracy,
but only in regions where the cusp--kink mixture produces an observable
distortion in the LISA band. Fourth, posterior anisotropy and reconstruction
quality are distinct: a large $\kappa_{\rm deg}$ can describe either a broad
degeneracy or a compact correlated measurement, depending on the covariance
area and marginalized widths. Finally, reconstruction accuracy must be assessed
separately from precision, since compact posteriors can still be biased.

These results emphasize that the inference problem for cosmic superstrings is
richer than a sensitivity forecast. A detected stochastic background may reveal
the presence of a cosmic-string-like source while leaving the microscopic origin
ambiguous. Conversely, when the spectrum contains measurable shape information,
Bayesian model comparison and posterior geometry can identify where LISA is
sensitive not only to the existence of a superstring background, but also to the
physical role of the intercommutation probability. This provides a concrete
framework for moving from detection toward reconstruction and model
discrimination for cosmic superstrings in the millihertz band.
 \section*{Acknowledgments}
%The work of SD is supported by the National Natural Science Foundation of China (NNSFC)
%under grant No. 12150610460. 
The work of RS is supported by the research project TAsP (Theoretical Astroparticle Physics) funded by the Istituto Nazionale di Fisica Nucleare (INFN).
\bibliography{bibliography}
\end{document}